\documentclass{ws-rv975x65}
\usepackage{ws-rv-van}             % numbered citation/references (default)
\makeindex
\newcommand{\Eq}[1]{Eq.~(\ref{#1})}
\newcommand{\Eqs}[2]{Eqs.(\ref{#1},\ref{#2})}

\newcommand{\ur}[1]{(\ref{#1})}

\newcommand{\beq}{\begin{equation}}
\newcommand{\eeq}{\end{equation}}

\newcommand{\la}[1]{\label{#1}}
\newcommand{\bea}{\begin{eqnarray}}
\newcommand{\eea}{\end{eqnarray}}
\newcommand{\ba}{\begin{array}}
\newcommand{\ea}{\end{array}}

\newcommand{\half}{{\textstyle{\frac{1}{2}}}}
\newcommand{\noi}{\noindent}

\newcommand{\n}{\nonumber}
\newcommand{\Tr}{{\rm Tr}}

\newcommand{\ECL}{E$\chi$L$\;$}
\newcommand{\at}{\overline{10}}
\newcommand{\oct}{\left({\bf 8},\frac{1}{2}\right)}
\newcommand{\dec}{\left({\bf 10},\frac{3}{2}\right)}
\newcommand{\adec}{\left({\bf\at},\frac{1}{2}\right)}
\newcommand{\pletthree}{\left({\bf 27},\frac{3}{2}\right)}
\newcommand{\pletone}{\left({\bf 27},\frac{1}{2}\right)}

\def\Dirac#1{#1\hskip-5pt/}
\def\dd{\Dirac\partial}

\def\Journal#1#2#3#4{{#1} {\bf #2}, #3 (#4)}
\def\NPA{{\em Nucl. Phys.} A}
\def\NPB{{\em Nucl. Phys.} B}
\def\PLB{{\em Phys. Lett.} B}
\def\PRL{\em Phys. Rev. Lett.}
\def\PRD{{\em Phys. Rev.} D}
\def\ZPA{{\em Z. Phys.} A}

\begin{document}

\chapter{Exotic baryon resonances in the Skyrme model\label{ch1}}

\author[Dmitri Diakonov and Victor Petrov]{Dmitri Diakonov and Victor Petrov}
%\index[aindx]{Author, F.} % or \aindx{Author, F.}
%\index[aindx]{Author, S.} % or \aindx{Author, S.}

\address{Petersburg Nuclear Physics Institute,\\
Gatchina, 188300, St. Petersburg, Russia\\
diakonov@thd.pnpi.spb.ru, victorp@thd.pnpi.spb.ru}

\begin{abstract}
We outline how one can understand the Skyrme model from the modern perspective.
We review the quantization of the $SU(3)$ rotations of the Skyrmion, leading
to the exotic baryons that cannot be made of three quarks. It is shown that in
the limit of large number of colours the lowest-mass exotic baryons can be studied
from the kaon-Skyrmion scattering amplitudes, an approach known after Callan and Klebanov.
We follow this approach and find, both analytically and numerically, a strong
$\Theta^+$ resonance in the scattering amplitude that is traced to the rotational
mode. The Skyrme model {\em does} predict an exotic resonance $\Theta^+$ but grossly
overestimates the width. To understand better the factors affecting the width,
it is computed by several methods giving, however, identical results.
In particular, we show that insofar as the width is small, it can be found from the transition
axial constant. The physics leading to a narrow $\Theta^+$ resonance is briefly reviewed and affirmed.
\end{abstract}

\body

\section{How to understand the Skyrme model}

It is astounding that Skyrme had suggested his model~\cite{Skyrme} as early as in 1961
before it has been generally accepted that pions are (pseudo) Goldstone bosons
associated with the spontaneous breaking of chiral symmetry, and of course
long before Quantum Chromodynamics (QCD) has been put forward as the microscopic
theory of strong interactions.

The revival of the Skyrme idea in 1983 is due to Witten~\cite{Witten}
who explained the {\it raison d'\^etre} of the Skyrme model from the viewpoint
of QCD. In the chiral limit when the light quark masses $m_u,m_d,m_s$ tend to zero,
such that the octet of the pseudoscalar mesons $\pi,K,\eta$ become nearly massless
(pseudo) Goldstone bosons, they are the lightest degrees of freedom of QCD.
The effective chiral Lagrangian (\ECL$\!$) for pseudoscalar mesons, understood
as an infinite expansion in the derivatives of the pseudoscalar (or chiral) fields,
encodes, in principle, full information about QCD. The famous two-term Skyrme
Lagrangian can be understood as a low-energy truncation of this infinite series.
Witten has added an important four-derivative Wess--Zumino term~\cite{WZ} to the original
Skyrme Lagrangian and pointed out that the overall coefficient in front of the \ECL
is proportional to the number of quark colours $N_c$.

Probably most important, Witten has shown that Skyrme's original idea of
getting the nucleon as a soliton of the \ECL  is justified in the limit of
large $N_c$ (since quantum corrections to a classical saddle point die out as
$1/N_c$) and that the `Skyrmion' gets correct quantum numbers upon quantization
of its rotations in ordinary and flavour spaces. Namely, if one restricts oneself
to two light flavours $u,d$, the lowest rotational states of a Skyrmion are
the nucleon with spin $J=\half$ and isospin $T=\half$ and the $\Delta$ resonance
with $J=\frac{3}{2}$ and $T=\frac{3}{2}$. For three light flavours $u,d,s$
the lowest rotational state is the $SU(3)$ octet with spin $\half$ and the
next is the decuplet with spin $\frac{3}{2}$, in full accordance with reality.
The statement appeared in Witten's `note added in proof' without a derivation
but a number of authors~\cite{rot-quantization} have derived the result
(it is reproduced in Section 2). Almost all of those authors noticed that
formally the next rotational excitation of the Skyrmion is an exotic  baryon
{\em anti}decuplet, again with spin $\half$, however few took it seriously.
It was only after the publication of Ref.~\cite{DPP-97} where it was predicted that
the lightest member of the antidecuplet, the $\Theta^+$ baryon, must be light
and narrow, that a considerable experimental and theoretical interest in the exotic
baryons has been aroused.

Soon after Witten's work it has been realized that it is possible to bring
the Skyrme model and the Skyrmion even closer to QCD and to the more customary language
of constituent quarks. It has been first noticed~\cite{DE,Dhar,DP-86} that
a simple chiral-invariant Lagrangian for massive (constituent) quarks $Q$ interacting
with the octet chiral field $\pi^A (A=1,...,8)$,
\beq
{\cal L}=\bar Q\left(\dd - M\,e^{\frac{i\pi^A\lambda^A\gamma_5}{F_\pi}}\right)Q,
\qquad \pi^A=\pi,K,\eta,
\la{L}\eeq
induces, via a quark loop in the external pseudoscalar fields (see Fig.~1), the \ECL whose
lowest-derivative terms coincide with the Skyrme Lagrangian, including automatically
the Wess--Zumino term, with the correct coefficient!

%%%%%%%%%%%%%%
%% FIGURE 1 %%
%%%%%%%%%%%%%%
\begin{figure}[]
\centerline{\epsfxsize=8.5cm\epsfbox{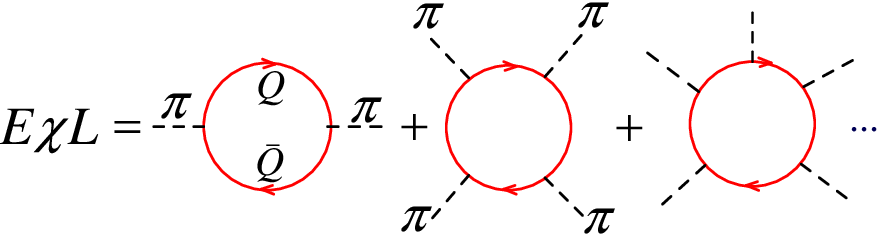}}
\caption{The effective chiral lagrangian (\ECL$\!\!$) is the quark loop in the external chiral field,
or the determinant of the Dirac operator \ur{L}. Its real part is the
kinetic energy term for pions, the Skyrme term and, generally, an infinite series
in derivatives of the chiral field. Its imaginary part is the Wess--Zumino term,
plus also an infinite series in derivatives~\cite{DE,Dhar,DPP-88}.}
\end{figure}

A step in the same direction, namely in bringing the Skyrme model closer
to the language of quarks, has been made in the chiral bag model by Brown,
Rho and collaborators~\cite{BrownRho}, for a review see Ref.~\cite{HosakaToki}

In fact, \Eq{L} can be derived in the instanton liquid model for the spontaneous
chiral symmetry breaking~\cite{DP-86} where a dynamical momentum-dependent quark
mass $M(p)$ is generated as an originally massless quark propagates through the
random ensemble of instantons and anti-instantons, each time flipping its helicity.
The low-energy quark Lagrangian \ur{L} is generally speaking nonlocal
which provides a natural ultraviolet cutoff. At low momenta, however, one can
treat the dynamical mass as a constant $M(0)\approx 350\,{\rm MeV}$~\cite{DP-86}.

It is implied that all gluon degrees of freedom, perturbative and not, are integrated
out when one comes to the effective low-energy quark Lagrangian of the type given by
\Eq{L}. Important, one does not need to add explicitly, say, the kinetic energy term
for pions to \Eq{L} (as several authors have originally suggested~\cite{MG,KRS,BB})
since the pion is a $Q\bar Q$ state itself and it propagates through quark loops,
as exhibited in the first graph in Fig.~1.

Understanding the quark origin of the \ECL it becomes possible to formulate
what is the Skyrmion in terms of quarks and demystify the famous prescription
of the Skyrme model that a chiral soliton with a topological (or winding) number
equal to unity, is in fact a fermion.

To that end, one looks for a trial chiral field capable of binding constituent
quarks. Let there be such a field $\pi({\bf x})$ that creates a bound-state
level for ``valence'' quarks, $E_{\rm val}$. Actually, one can put $N_c$ quarks at
that level in the antisymmetric colour state, as the chiral field is colour-blind.
The energy penalty for creating the trial field is given by the same Lagrangian \ur{L}.
It is the aggregate energy of the negative-energy Dirac sea of quarks distorted
by the trial field, $E_{\rm sea}$; it should be also multiplied by $N_c$
since all negative-energy levels should be occupied and they are $N_c$-fold
degenerate in colour. Therefore, the full energy of a state with baryon number unity
and made of $N_c$ quarks, is a sum of two functionals~\cite{DP8,DPP-88},
\beq
{\cal M}_N = N_c\left(E_{\rm val}[\pi({\bf x})]+E_{\rm sea}[\pi({\bf x})]\right).
\la{MN}\eeq
Schematically it is shown in Fig.~2. The self-consistent (or mean) pion field binding
quarks is the one minimizing the nucleon mass. Quantum fluctuations about it are
suppressed insofar as $N_c$ is large. The condition that the winding number of
the trial field is unity needs to be imposed to get a deeply bound state, that is
to guarantee that the baryon number is unity~\cite{DPP-88}. The Skyrmion is, thus,
nothing but the {\bf mean chiral field binding quarks in a baryon}.

%%%%%%%%%%%%%%%
%% FIGURE 2 %%
%%%%%%%%%%%%%%%
\begin{figure}[]
\centerline{\epsfxsize=11cm\epsfbox{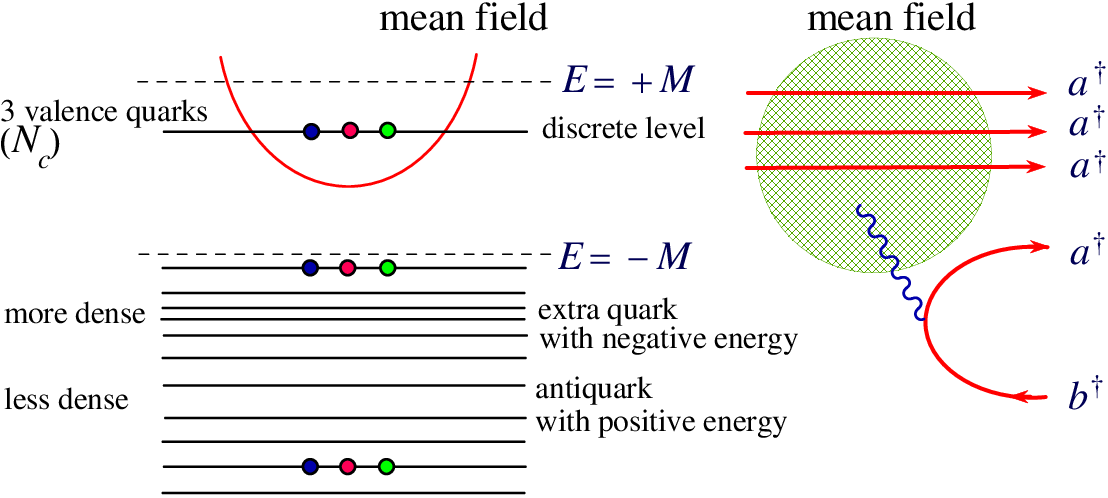}}
\caption{{\it Left:} If the trial pion field is large enough (shown schematically by
the solid curve), there is a discrete bound-state level for $N_c$ `valence'
quarks, $E_{\rm val}$. One has also to fill in the negative-energy
Dirac sea of quarks (in the absence of the trial pion field it corresponds
to the vacuum). The spectrum of the negative-energy levels is shifted in
the trial pion field, its aggregate energy, as compared to the free case,
being $E_{\rm sea}$.
{\it Right:} Equivalent view of baryons, where the polarized Dirac sea is presented
as $Q\bar Q$ pairs.
}
\end{figure}

This model of baryons, called the {\it Chiral Quark Soliton Model} or the
{\it Relativistic Mean Field Approximation}, apparently interpolates between
the nonrelativistic constituent quark model and the Skyrme model, making sense
and showing the limitations of both. Indeed, if the mean field happens to be small,
the valence level is shallow, the Dirac sea is weakly distorted, and there are
few antiquarks. In this case the model reproduces the well-known spin and space
quark wave functions of the nonrelativistic models for baryons~\cite{DP-05}.
If, on the contrary, the mean field happens to be very broad, the valence level
dives towards the negative-energy Dirac sea, and there are ${\cal O}(N_c)$ (that is many!)
additional $Q\bar Q$ pairs in a baryon, whose energy can be approximated
by the two- and four-derivative Skyrme Lagrangian. The realistic picture
is somewhere in between the two extremes.

Decoding the Skyrme model in terms of quarks allows one to answer important
questions that cannot even be asked in the Skyrme model. For example,
one can find out parton distributions in nucleons, satisfying all
general sum rules and positivity constraints~\cite{DPPPW}, the light-cone
distribution amplitudes~\cite{PP-03} or, {\it e.g.} the wave functions
of the 5-quark components in nucleons~\cite{DP-05}. For reviews of the
model see~\cite{Ripka,DP-00}. \\

To summarize this introduction: The original Skyrme's idea is well founded
from the modern QCD viewpoint. There is no mystics in the identification of
the pion field winding number with the baryon number, and in the Skyrmion
being a fermion (at odd $N_c$). The chiral soliton field, the Skyrmion,
is nothing but the self-consistent mean field binding $N_c$ valence quarks
and distorting the Dirac sea such that additional $Q\bar Q$ pairs are
necessary present in a baryon.

At the same time, one cannot expect a fully quantitative description of reality
in the concrete two-terms Skyrme's original model as an infinite series
in the derivatives in the \ECL is truncated: it is similar to replacing $e^{-x}$
by $1\!-\!x$. What is even worse, there are no explicit valence quarks in
the Skyrme model as they cannot be separated from the sea.

In what follows, we shall nevertheless mainly deal with the concrete model by Skyrme
(supplemented by the Wess--Zumino term) in order to study certain
qualitative features of the exotic baryon resonances, {\it i.e.} those that
by quantum numbers cannot be composed of three quarks only but need additional
$Q\bar Q$ pairs.

\section{Rotational states of the $SU(3)$ Skyrmion}

The results of this section are general in the sense that they are independent
on whether one takes literally the Skyrme model or a more sophisticated chiral quark model.

The standard choice of the saddle point field is the ``upper-left corner hedgehog'' Ansatz:
\beq
U_0({\bf x})\equiv e^{i\pi_0^A({\bf x})\lambda^A}
= \left(\begin{array}{cc}
e^{i(\mbox{\boldmath{$n$}}
\cdot\mbox{\boldmath{$\tau$}}) P(r)} &
\begin{array}{c} 0\\ 0\end{array}\\ \begin{array}{cc}0\;\;\; &\;\;\;0 \end{array} & 1
\end{array}\right),\qquad {\bf n}=\frac{{\bf x}}{r},
\la{hedgehog}\eeq
where $\lambda^A$ are eight Gell-Mann matrices, $\tau^i$ are three Pauli matrices,
and $P(r)$ is a spherically symmetric function called the profile of the Skyrmion.
In the chiral limit $m_u=m_d=m_s=0$ any $SU(3)$ rotation of the saddle point field,
$R U_0 R^\dagger,\;R\in SU(3)$, is also a saddle point. We consider a slowly rotating
Ansatz,
\beq
U({\bf x},t)=R(t)U_0({\bf x})R^\dagger(t)
\la{rotAnsatz}\eeq
and plug it into the \ECL. The degeneracy of the saddle point in the flavour rotations
means that the action will not depend on $R$ itself but only on the time derivatives
$\dot R$. We do not consider the rotation angles as small but rather expand the action
in angular velocities. In fact, one has to distinguish between the `right' ($\Omega_A$)
and `left' ($\omega_A$) angular velocities defined as
\beq
\Omega_A=-i\Tr(R^\dagger \dot R\lambda^A),\qquad
\omega_A=-i\Tr(\dot R R^\dagger\lambda^A),\qquad
\Omega^2=\omega^2=2\Tr\dot R^\dagger\dot R.
\la{angvel}\eeq

Given the Ansatz \ur{hedgehog} one expects on symmetry grounds the following Lagrangian
for slow rotations:
\beq
{\cal L}_{\rm rot} =
\frac{I_1}{2}\left(\Omega_1^2+\Omega_2^2+\Omega_3^2\right)
+ \frac{I_2}{2}\left(\Omega_4^2+\Omega_5^2 +\Omega_6^2+\Omega_7^2\right)
- \frac{N_cB}{2\sqrt{3}}\Omega_8
\la{Lrot}\eeq
where $I_{1,2}$ are the two soliton moments of inertia that are functionals
of the profile function $P(r)$. Rotation along the 8th axis in flavour space,
$R=\exp(i\alpha_8\lambda^8)$, commutes with the `upper-left-corner' Ansatz,
therefore there is no quadratic term in $\Omega_8$. However there is a
Wess--Zumino term resulting in a term linear in $\Omega_8$ proportional
to the baryon number $B$. In the chiral quark models this term arises from
the extra bound-state levels for quarks~\cite{Blotz}.

To quantize this rotational Lagrangian one uses the canonical
quantization procedure. Namely, one introduces eight `right' angular momenta
$J_A$ canonically conjugate to `right' angular velocities $\Omega_A$,
\beq J_A=-\frac{\partial {\cal L}_{\rm rot}}{\partial \Omega_A},
\la{JA}\eeq
and writes the rotational Hamiltonian as
\beq
{\cal H}_{\rm rot}=\Omega_A J_A-{\cal L}_{\rm rot}
=\frac{J_1^2+J_2^2+J_3^2}{2I_1}+\frac{J_4^2+J_5^2+J_6^2+J_7^2}{2I_2}
\la{Hrot}\eeq
with the additional quantization prescription following from \Eq{JA},
\beq
J_8=\frac{N_cB}{2\surd{3}}.
\la{quantiz}\eeq
The quantization amounts to replacing classical angular momenta $J_A$
by $SU(3)$ generators satisfying the $su(3)$ algebra: $[J_AJ_B]=if_{ABC}J_C$
where $f_{ABC}$ are the $su(3)$ structure constants. These generators act
on the matrix $R$ on the right, $\exp(i\alpha^AJ_A))R\exp(i(-\alpha_AJ_A))
=R\exp(-i\alpha^A\lambda^A/2)$. For the first three generators ($A=1,2,3$)
this is equivalent, thanks to the hedgehog Ansatz \ur{hedgehog}, to rotating the
space axes $x,y,z$. Therefore, $J_{1,2,3}$ are in fact spin generators.
%The rest 5 generators do not have an obvious physical meaning. The

One can also introduce `left' angular momenta $T_A$ canonically conjugate
to the `left' angular velocities $\omega_A$; they satisfy the same $su(3)$ algebra,
$[T_AT_B]=if_{ABC}T_C$, whereas $[T_A J_B]=0$. These generators act
on the matrix $R$ on the left, $\exp(i\alpha^AT_A))R\exp(i(-\alpha_AT_A))
=\exp(i\alpha^A\lambda^A/2)R$, and hence have the meaning of $SU(3)$ flavour
generators. The quadratic Casimir operator can be written using either
`left' or `right' generators as
\beq
J_AJ_A =T_AT_A =  C_2(p,q) = \frac{1}{3}\left[p^2+q^2+pq+3(p+q)\right]
\la{C2}\eeq
where $C_2(p,q)$ is the eigenvalue of the quadratic Casimir operator
for an irreducible representation $r$ of $SU(3)$, labeled by two integers $(p,q)$.
The rotational wave functions of chiral soliton are thus finite $SU(3)$ rotation
matrices $D^r_{T,T_3,Y;J,J_3,Y'}(R)$ characterized by the eigenvalues of
the commuting generators. For the $SU(2)$ group they are called
Wigner finite-rotation matrices and depend on 3 Euler angles; in $SU(3)$
there are 8 `Euler' angles. The general rotational functions (with important sign
subtleties) are given in the Appendix of Ref.~\cite{Blotz}, and practically
useful examples are given explicitly in Ref.~\cite{DP-05}.

One can visualize the rotational wave functions as a product of two same $SU(3)$
weight diagrams: one for the eigenvalues of the flavour (`left') generators, and
the other for the eigenvalues of `right' generators including the spin. Important,
the quantization condition \ur{quantiz} means that not all $SU(3)$ representations
can be viewed as rotational states of a Skyrmion. Taking baryon number $B\!=\!1$
and $N_c=3$ and recalling that $J_8=Y'\sqrt{3}/2$ where $Y$ is the hypercharge,
the condition \ur{quantiz} means that only those multiplets are rotational states
that contain particles with $Y'=1$ or, more generally,
\beq
Y'=\frac{N_c}{3}.
\la{Y}\eeq
The lowest $SU(3)$ multiplets meeting this condition are the octet, the decuplet
and the antidecuplet, see Fig.~3.

%%%%%%%%%%%%%%%
%% FIGURE 3 %%
%%%%%%%%%%%%%%%
\begin{figure}[]
\centerline{\epsfxsize=5cm\epsfbox{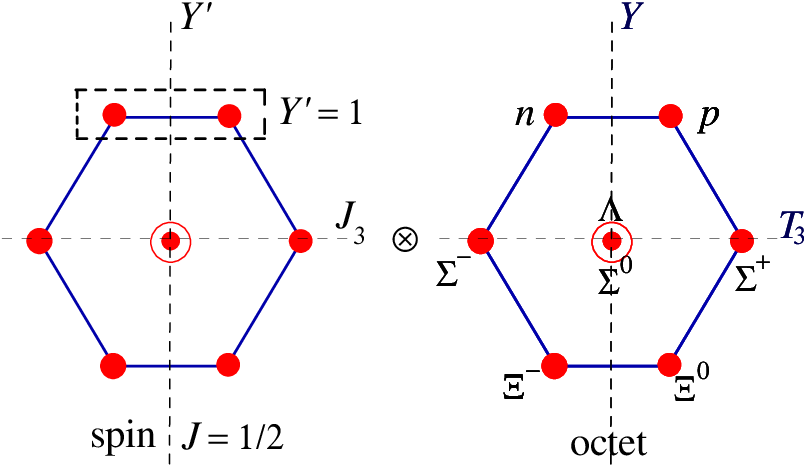}}\hspace{1cm}
{\epsfxsize=5.5cm\epsfbox{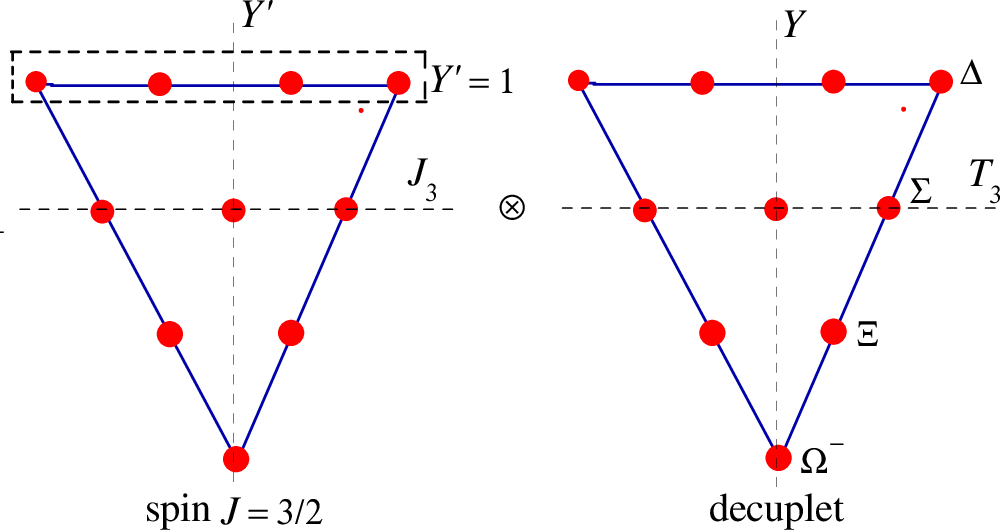}}\hspace{1cm}
{\epsfxsize=5cm\epsfbox{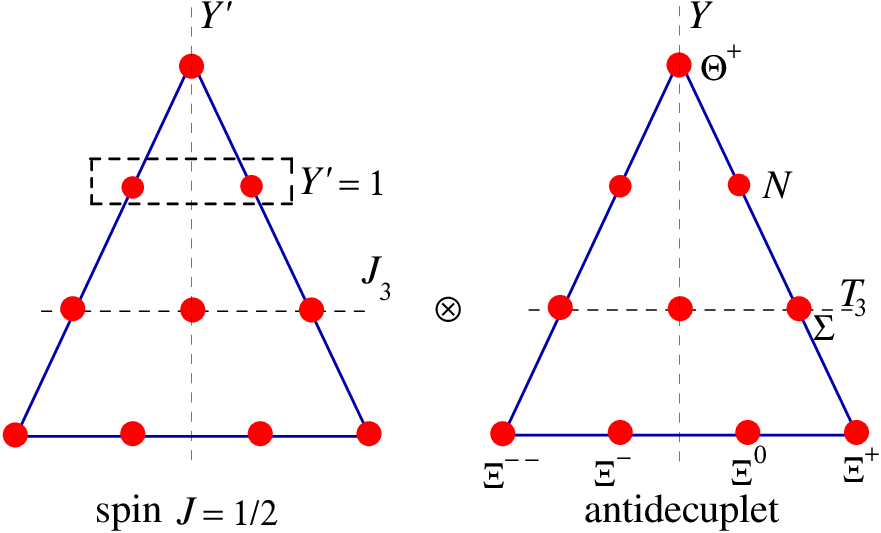}}
\caption{The lowest rotational states of a Skyrmion, satisfying the condition $Y'=1$:
$\oct$, $\dec$, $\adec$.
The number of states with $Y'=1$, if one equates it to $2J+1$, determines the spin $J$
of the particles in the multiplet.
}
\end{figure}
It is remarkable that the lowest rotational states of the Skyrmion are exactly those
observed in nature. The third is the antidecuplet with spin $\half$. In the three vertices
of the big triangle shown in Fig.~3, bottom right, there are baryons that are explicitly
exotic, in the sense that they cannot be composed of three quarks but need an additional
quark-antiquark pair. For example, the $\Theta^+$ baryon at the top of the triangle
can be composed minimally of $uudd\bar s$ quarks, {\it i.e.} it is a {\em pentaquark}.
Seven baryons that are not in the vertices of the antidecuplet are cryptoexotic,
in the sense that their quantum numbers can be, in principle, arranged from three quarks,
however their expected properties are quite different from those of the similar members
of a baryon octet.

It should be remembered, however, that strictly speaking the whole Skyrmion approach
to baryons is justified in the limit of large $N_c$. Whether $N_c\!=\!3$ is ``large
enough'' is a question to which there is no unique answer: it depends on how large
are the $1/N_c$ corrections to a particular physical quantity. Therefore, one has
to be able to write equations with $N_c$ being a free parameter. In particular,
at arbitrary $N_c$ one has to construct explicitly $SU(3)$ flavor multiplets
that generalize the lightest baryon multiplets $\oct$, $\dec$, $\adec$, {\it etc.},
to arbitrary $N_c$. We do it in the next section following Ref.~\cite{DP-04}
that generalizes previous work on this subject~\cite{DuPrasz,Cohen}.

\section{Rotational multiplets at arbitrary $N_c$}

We remind the reader that a generic $SU(3)$ multiplet or irreducible representation
is uniquely determined by two non-negative integers $(p,q)$ having the meaning
of upper (lower) components of the irreducible $SU(3)$ tensor
$T^{\{f_1...f_p\}}_{\{g_1...g_q\}}$ symmetrized both in upper and lower
indices and with a contraction with any $\delta^{g_n}_{f_m}$ being zero.
Schematically, $q$ is the number of boxes in the lower line of the
Young tableau depicting an $SU(3)$  representation and $p$ is the
number of extra boxes in its upper line, see Fig.~4.

%%%%%%%%%%%%%%%
%% FIGURE 4  %%
%%%%%%%%%%%%%%%
\begin{figure}[]
\centerline{\epsfxsize=9cm\epsfbox{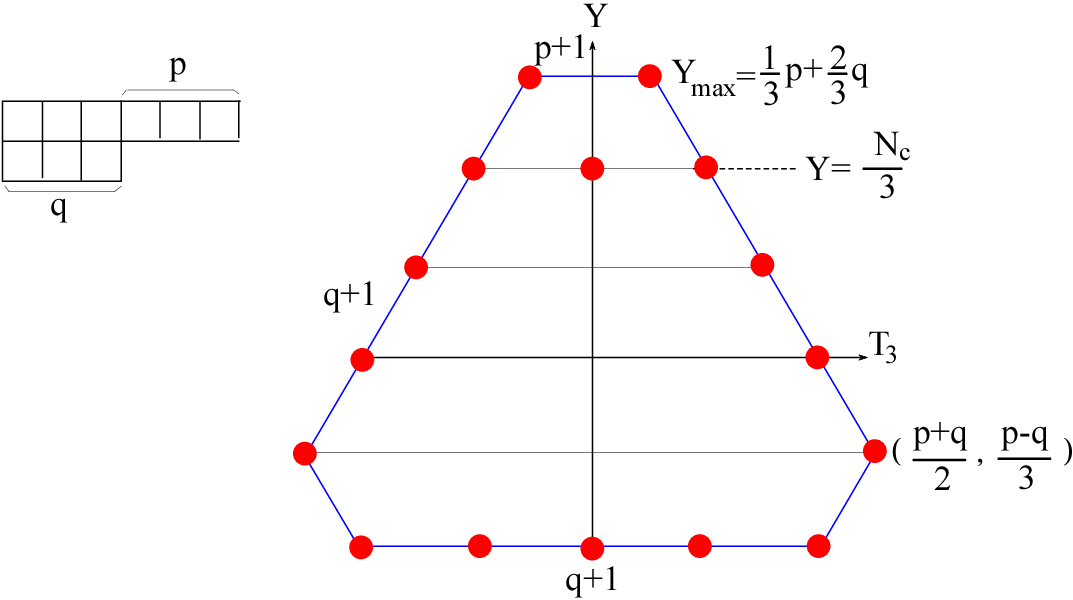}}
\caption{A generic $SU(3)$ multiplet is, on the one hand, defined by the
Young tableau and on the other hand can be characterized by quantum numbers
$(T_3,Y)$ of its members filling a hexagon in the $(T_3,Y)$ axes (the weight
diagram).
}
\end{figure}
\noi

The dimension of a representation or the number of particles in the multiplet is
\beq
{\rm Dim}(p,q)=(p+1)(q+1)\left(1+\frac{p+q}{2}\right).
\la{Dim}\eeq
On the weight $(T_3,Y)$ diagram where $T_3$ is the third projection
of the isospin and $Y$ is the hypercharge, a generic
$SU(3)$ representation is depicted by a hexagon, whose upper
horizontal side contains $p+1$ `dots' or particles, the adjacent
sides contain $q+1$ particles, with alternating $p+1$ and $q+1$
particles in the rest sides, the corners included --  see Fig.~4.
If either $p$ or $q$ is zero, the hexagon reduces to a triangle.
Particles on the upper (horizontal) side of the hexagon have
the hypercharge
\beq
Y_{\rm max}=\frac{1}{3}\,p+\frac{2}{3}\,q
\la{Ymax}\eeq
being the maximal possible hypercharge of a multiplet with given $(p,q)$.
Given that
\beq
\sum_{A=1}^3J_A^2=J(J+1),\qquad
\sum_{A=1}^8J_A^2=C_2(p,q),\qquad J_8^2=\frac{N_c^2}{12},
\la{sums}\eeq
one gets from \Eq{Hrot} the rotational energy of baryons with given spin $J$
and belonging to representation $(p,q)$:
\beq
{\cal E}_{\rm rot}(p,q,J)=
\frac{C_2(p,q)-J(J+1)-\frac{N_c^2}{12}}{2I_2}+\frac{J(J+1)}{2I_1}.
\la{Erot}\eeq
Only those multiplets are realized as rotational excitations that have
members with hypercharge $Y=\frac{N_c}{3}$; if the number of particles with
this hypercharge is $n$ the spin $J$ of the multiplet is such that $2J+1=n$.
It is easily seen that the number of particles with a given $Y$
is $\frac{4}{3}p+\frac{2}{3}q+1-Y$ and hence the spin of the
allowed multiplet is
\beq
J= \frac{1}{6}(4p+2q-N_c).
\la{J}\eeq
A common mass ${\cal M}_0$ must be added to \Eq{Erot} to
get the mass of a particular multiplet. Throughout this section we disregard
the splittings inside multiplets as due to nonzero current strange quark mass.

The condition that a horizontal line $Y=\frac{N_c}{3}$ must be inside
the weight diagram for the allowed multiplet leads to the requirement
\beq
\frac{N_c}{3}\leq Y_{\rm max} \qquad {\rm or} \quad p+2q\geq N_c
\la{cond}\eeq
showing that at large $N_c$ multiplets must have a high dimension!

We introduce a non-negative number which we name ``exoticness'' $X$ of a
multiplet defined as~\cite{DP-04}
\beq
Y_{\rm max}\;=\;\frac{1}{3}p+\frac{2}{3}q\;\equiv\; \frac{N_c}{3}+X, \qquad
X\geq 0.
\la{X}\eeq
Combining \Eqs{J}{X} we express $(p,q)$ through $(J,X)$:
\bea
\n
p &=& 2 J - X, \\
\la{pq_JX} q &=& \frac{1}{2}N_c +2X-J.
\eea
The total number of
boxes in Young tableau is $2q+p=N_c+3X$. Since we are dealing with
unity baryon number states, the number of quarks in the multiplets
we discuss is $N_c$, {\em plus} some number of quark-antiquark
pairs. In the Young tableau, quarks are presented by single boxes
and antiquarks by double boxes. It explains the name
``exoticness":  $X$ gives the minimal number of additional quark-antiquark
pairs one needs to add on top of the usual $N_c$ quarks to compose
a multiplet.

Putting $(p,q)$ from \Eq{pq_JX} into \Eq{Erot} we obtain the rotational
energy of a soliton as function of the spin and exoticness of the multiplet:
\beq
{\cal E}_{\rm rot}(J,X)=\frac{X^2+X(\frac{N_c}{2}+1-J)+\frac{N_c}{2}}{2I_2}
+\frac{J(J+1)}{2I_1}.
\la{ErotJX}\eeq
We see that for given $J\leq \frac{N_c}{2}+1$ the multiplet mass is a
monotonically growing function of $X$: the minimal-mass multiplet has
$X=0$. Masses of multiplets with increasing exoticness are:
\bea
\la{X0}
{\cal M}_{{\rm X}\!=\!0}(J) &=&
{\cal M}_0^\prime+\frac{J(J+1)}{2I_1},\qquad {\rm where}\quad
{\cal M}_0^\prime\equiv {\cal M}_0+\frac{N_c}{4I_2}, \\
\n\\
\la{X1}
{\cal M}_{{\rm X}\!=\!1}(J) &=&
{\cal M}_0^\prime+\frac{J(J+1)}{2I_1}+1\cdot \frac{\frac{N_c}{2}+2-J}{2I_2}, \\
\n\\
\la{X2}
{\cal M}_{{\rm X}\!=\!2}(J) &=&
{\cal M}_0^\prime+\frac{J(J+1)}{2I_1}+2\cdot
\frac{\frac{N_c}{2}+2-J}{2I_2}+\frac{1}{I_2},
\qquad {\rm etc.}
\eea
%%%%%%%%%%%%%%%
%% FIGURE 5 %%
%%%%%%%%%%%%%%%
\begin{figure}[t]
\centerline{\epsfxsize=9cm\epsfbox{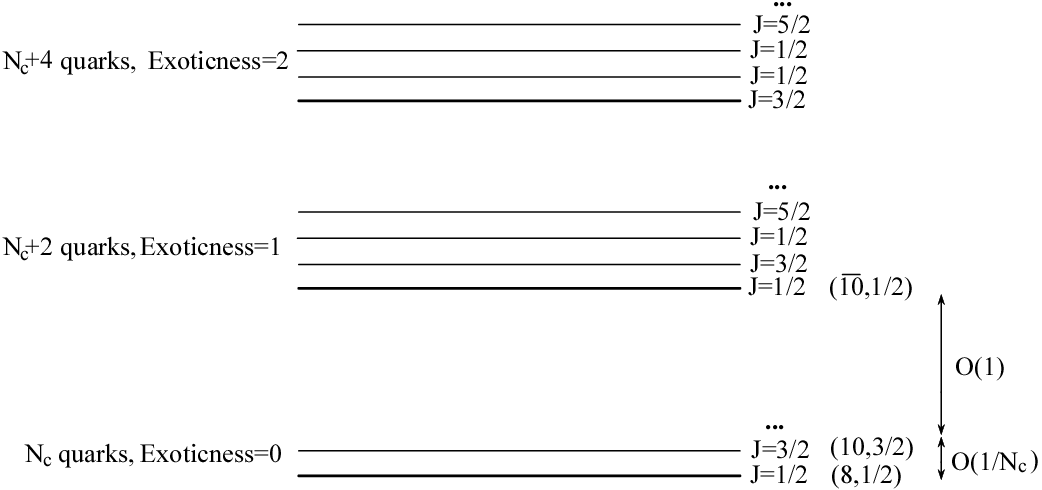}}
\caption{Rotational excitations form a sequence of bands.
}
\end{figure}

At this point it should be recalled that both moments of inertia
$I_{1,2}=O(N_c)$, as is ${\cal M}_0$. We see from Eqs.\ur{X0}-\ur{X2}
that multiplets fall into a sequence of rotational bands each labeled
by its exoticness with small $O(1/N_c)$ splittings inside the bands.
The separation between bands with different exoticness is $O(1)$.
The corresponding masses are schematically shown in Fig.~5.

The lowest band is non-exotic ($X\!=\!0$); the multiplets are
determined by $(p,q)=\left(2J,\frac{N_c}{2}-J\right)$, and their
dimension is $\;{\rm Dim}=(2J+1)(N_c+2-2J)(N_c+4+2J)/8$ which in the
particular (but interesting) case of $N_c=3$ becomes 8 for spin
one half and 10 for spin 3/2. These are the correct lowest
multiplets in real world, and the above multiplets are their
generalization to arbitrary values of $N_c$. To make baryons
fermions one needs to consider only odd $N_c$.
%%%%%%%%%%%%%%%
%% FIGURE 6  %%
%%%%%%%%%%%%%%%
\begin{figure}[b]
\centerline{\epsfxsize=9cm\epsfbox{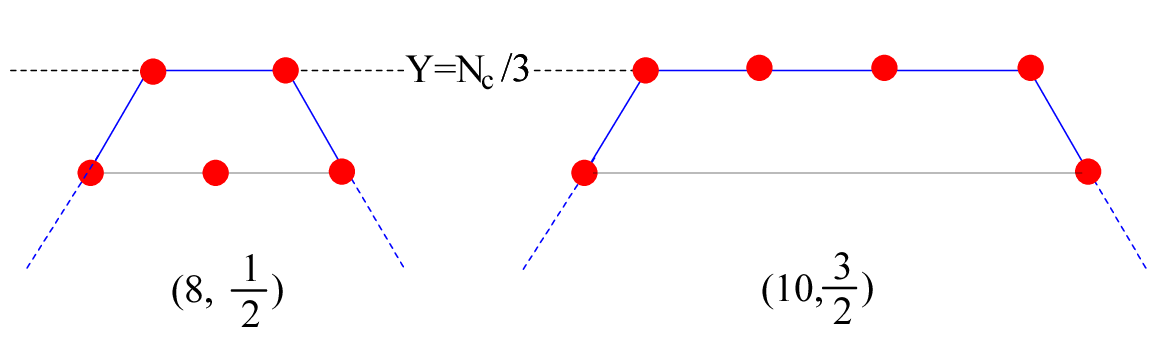}}
\caption{Non-exotic ($X\!=\!0$) multiplets that can be composed of $N_c$
quarks.}
\end{figure}

Recalling that $u,d,s$ quarks' hypercharges are 1/3, 1/3 and -2/3,
respectively, one observes that all baryons of the non-exotic
$X\!=\!0$ band can be made of $N_c$ quarks. The upper side of
their weight diagrams (see Fig.~6) is composed of $u,d$ quarks
only; in the lower lines one consequently replaces $u,d$ quarks by
the $s$ one. This is how the real-world $\oct$ and $\dec$
multiplets are arranged and this property is preserved in their
higher-$N_c$ generalizations. The construction coincides with that
of Ref.~\cite{DuPrasz}. At high $N_c$ there are further multiplets
with spin 5/2 and so on. The maximal possible spin at given $N_c$
is $J_{\rm max}=\frac{N_c}{2}$: if one attempts higher spin, $q$
becomes negative.

%%%%%%%%%%%%%%%
%% FIGURE 7 %%
%%%%%%%%%%%%%%%
\begin{figure}[t]
\centerline{\epsfxsize=6.5cm\epsfbox{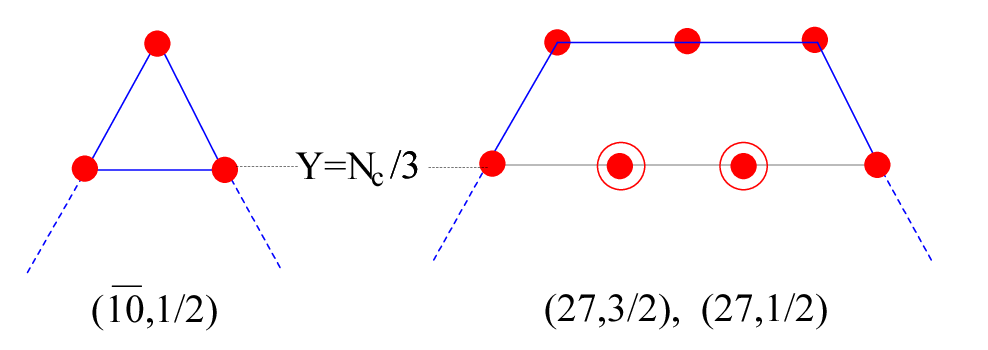}\hspace{1cm}\epsfxsize=4.5cm\epsfbox{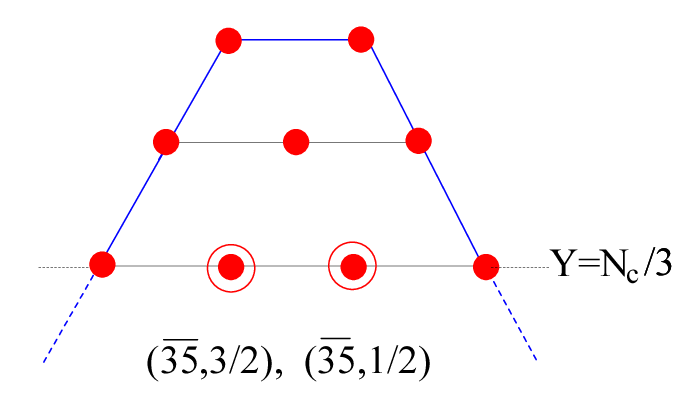}}
\caption{Exotic ($X\!=\!1$) multiplets (left and middle graphs) that can be composed of $N_c$ quarks
and one extra $Q\bar Q$ pair. An example of an $X\!=\!2$ multiplet that can be composed with two additional
$Q\bar Q$ pairs, is shown on the right.}
\end{figure}

The rotational bands for  $X\!=\!1$ multiplets are shown in Fig.~7, left
and middle graphs. The upper side of the weight diagram is exactly one unit higher
than the line $Y=\frac{N_c}{3}$ which is non-exotic, in the sense that its
quantum numbers can be, in principle, achieved from exactly $N_c$ quarks.
However, particles corresponding to the upper side of the weight diagram
cannot be composed of $N_c$ quarks but require at least one additional
$\bar s$ quark and hence {\em one additional quark-antiquark pair} on top of
$N_c$ quarks.

The multiplet shown in Fig.~7, left, has only one particle with
$Y=\frac{N_c}{3}+1$. It is an isosinglet with spin $J\!=\!\frac{1}{2}$, and
in the quark language is built of $(N_c+1)/2$ $ud$ pairs and one $\bar s$
quark. It is the generalization of the $\Theta^+$ baryon to arbitrary odd
$N_c$. As seen from \Eqs{Dim}{pq_JX}, the multiplet to which the
``$\Theta^+$" belongs is characterized by $(p,q)=\left(0,(N_c+3)/2\right)$,
its dimension is $(N_c+5)(N_c+7)/8$ becoming the $\adec$ at $N_c\!=\!3$.
Its splitting with the $N_c$ generalization of the non-exotic $\oct$
multiplet follows from \Eq{X1}:
\beq
{\cal M}_{\at,\frac{1}{2}}-{\cal M}_{8,\frac{1}{2}}=\frac{N_c+3}{4I_2},
\la{antiten_oct}\eeq
a result first found in Ref. \cite{Cohen}. Here and in what
follows we denote baryon multiplets by their dimension at $N_c\!=\!3$
although at $N_c\!>\!3$ their dimension is higher, as given by \Eq{Dim}.

The second rotational state of the $X\!=\!1$ sequence has $J=\frac{3}{2}$;
it has $(p,q)=(2,(N_c+1)/2)$ and dimension $3(N_c+3)(N_c+9)/8$ reducing to
the multiplet $\pletthree$ at $N_c=3$, see Fig.~7, middle. In fact there
are two physically distinct multiplets there. Indeed, the weights in the
middle of the second line from top on the weight diagram with
$Y=\frac{N_c}{3}$ are twice degenerate, corresponding to spin 3/2 and 1/2.
Therefore, there is another  $3(N_c+3)(N_c+9)/8$-plet  with unit exoticness,
but with spin 1/2. At $N_c\!=\!3$ it reduces to $\pletone$. The splittings
with non-exotic multiplets are
\bea
\la{273}
{\cal M}_{27,\frac{3}{2}}-{\cal M}_{10,\frac{3}{2}}
&=&\frac{N_c+1}{4I_2},\\
\n\\
\la{271}
{\cal M}_{27,\frac{1}{2}}-{\cal M}_{8,\frac{1}{2}}
&=&\frac{N_c+7}{4I_2}.
\eea
The $X\!=\!1$ band continues to the maximal spin $J_{\rm max}=(N_c+4)/2$
where $q$ becomes zero.

The $X\!=\!2$ rotational band (see Fig.~7, right) starts from two states
with spin 3/2 and 1/2 both belonging to the $SU(3)$ representation
$(p,q,{\rm Dim})=\left(1,(N_c+5)/2,(N_c+7)(N_c+11)/4\right)$. It reduces to the
${\overline{\bf 35}}$ multiplet at $N_c\!=\!3$. Their splittings with
non-exotic multiplets are
\bea
\la{353}
{\cal M}_{\overline{35},\frac{3}{2}}-{\cal M}_{10,\frac{3}{2}}
&=&\frac{N_c+3}{2I_2},\\
\n\\
\la{351}
{\cal M}_{\overline{35},\frac{1}{2}}-{\cal M}_{8,\frac{1}{2}}
&=&\frac{N_c+6}{2I_2}.
\eea
The maximal spin of the $X\!=\!2$ rotational band is
$J_{\rm max}=(N_c+8)/2$.

The upper side in the weight diagram in Fig.~7, right, for the $X\!=\!2$ sequence
has hypercharge $Y_{\rm max}=\frac{N_c}{3}+2$. Therefore, one needs
{\em two} $\bar s$ quarks to get that hypercharge and hence the multiplets
can be minimally constructed of $N_c$ quarks plus {\em two additional
quark-antiquark pairs}.

Disregarding the rotation along the 1,2,3 axes (for example taking
only the lowest $J$ state from each band) we observe from
\Eq{ErotJX} that at large $N_c$ the spectrum is equidistant in
exoticness,
\beq
{\cal E}_{\rm rot}(X)=\frac{N_c(X+1)}{4I_2},
\la{EX}\eeq
with the spacing $\frac{N_c}{4I_2}=O(1)$. It is consistent with the fact explained
in the next Section, that at large $N_c$ the rotation corresponding to the
excitations of exoticness is actually a small-angle precession equivalent
to small oscillations whose quantization leads to an equidistant spectrum.
We stress that there is no deformation of the Skyrmion by rotation until $X$ becomes
of the order of $N_c$~\cite{DP-04}.
\Eq{EX} means that each time we add a quark-antiquark pair it costs at large $N_c$ the same
\beq
{\rm energy\;of\;a\;Q\bar Q\;pair} = \omega_{\rm rot}=\frac{N_c}{4I_2}=O(N_c^0).
 \la{cost}\eeq
Naively one may think that this quantity should be approximately
twice the constituent quark mass $M\approx 350\,{\rm MeV}$. Actually, it can be
much less than that. For example, an inspection of $I_2$ in the Chiral Quark Soliton
Model \cite{DPP-88,Blotz} shows that the pair energy is strictly less than $2M$;
in fact $1/I_2$ tends to zero in the limit when the baryon size blows up.

In physical terms, the energy cost of adding a $Q\bar Q$ pair can be small
if the pair is added in the form of a Goldstone boson. The energy penalty for making,
say, the $\Theta^+$ baryon from a nucleon would be exactly zero in the chiral
limit and were baryons infinitely large. In reality, one has to create a
pseudo-Goldstone K-meson and to confine it inside the baryon of the size $\geq 1/M$.
It costs roughly
\beq
m(\Theta)-m(N)\approx \sqrt{m_K^2+{\bf p}^2}\leq \sqrt{495^2+350^2}=606\,{\rm MeV}.
\la{cost}\eeq
Therefore, one should expect the exotic $\Theta^+$ around 1540 MeV where indeed
it has been detected in a number of experiments!

\section{Rotational wave functions}

It is helpful to realize how do the rotational wave functions $\Psi(R)$ look like
for various known (and unknown) baryons. To that end, one needs a concrete
parameterization of the $SU(3)$ rotation matrix $R$ by 8 `Euler' angles: the wave
functions are in fact functions of those angles.

In general, the parameter space of an $SU(N)$ group is a direct product of odd-dimensional
spheres, $S^3\times S^5\times\ldots\times S^{2N-1}$. For $SU(3)$, it is a product of the
spheres $S^3\times S^5$. A general $SU(3)$ matrix $R$ can be written as $R=S_3R_2$
where $R_2$ is a general $SU(2)$ matrix with three parameters, put in the upper-left
corner, and $S_3$ is an $SU(3)$ matrix of a special type with five parameters,
see Appendix A in Ref.~\cite{DP-05}.

To be specific, let us consider the rotational wave function corresponding to
the exotic $\Theta^+$ baryon. For general $N_c$ its (complex conjugate) wave
function is given by~\cite{DP-05}
\beq
\Theta_k(R)^*=\left(R^3_3\right)^{N_c-1}R^3_k\,
\la{rot_w_f_Theta}\eeq
where $k=1,2$ is the spin projection and $R^3_k$ is the $k^{\rm th}$ matrix element
in the $3^{\rm d}$ row of the $3\times 3$ matrix $R$. Using the concrete parameterization
of Ref.~\cite{DP-05} \Eq{rot_w_f_Theta} becomes
\beq
\Theta(R)^*\sim \left(\cos\theta\,\cos\phi\right)^{N_c},
\la{rot_w_f_Theta-1}\eeq
where $\theta,\phi\in(0,\frac{\pi}{2})$ are certain angles parameterizing the $S^5$
sphere; $\theta=\phi=0$ corresponds to the North pole of that sphere. We see that
although for $N_c\!=\!3$ the typical angles in the wave function are large
such that it is spread over both $S^3$ and $S^5$ globes, at $N_c\to\infty$ the wave
function is concentrated near the North pole of $S^5$ since
\beq
\theta\sim\phi\sim\sqrt{\frac{2}{N_c}}\stackrel{N_c\to\infty}{\longrightarrow} 0.
\la{North_pole}\eeq
This is illustrated in Fig.~8.

%%%%%%%%%%%%%%%
%% FIGURE 8  %%
%%%%%%%%%%%%%%%
\begin{figure}[]
\centerline{\epsfxsize=5cm\epsfbox{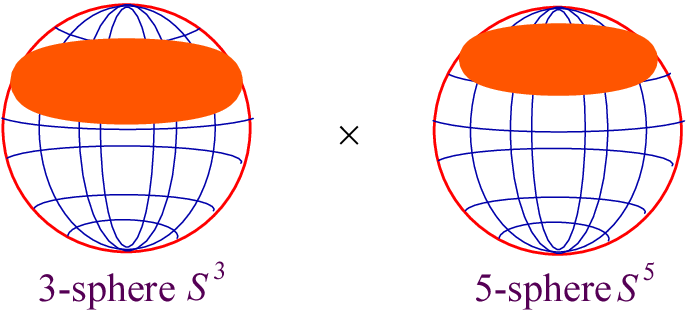}\hspace{2cm}
\epsfxsize=5cm\epsfbox{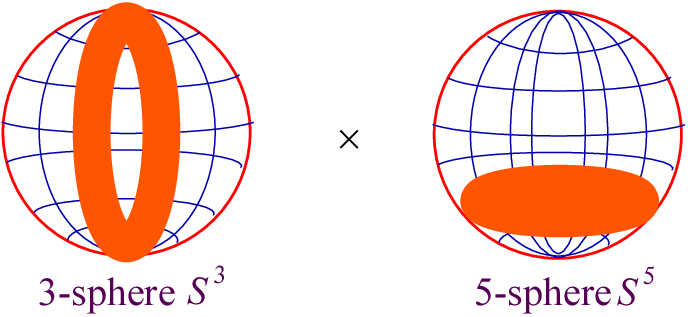}}
\vspace{1cm}
\centerline{\epsfxsize=5cm\epsfbox{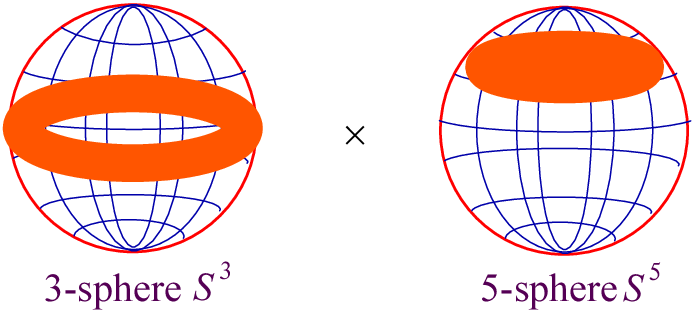}\hspace{2cm}
\epsfxsize=5cm\epsfbox{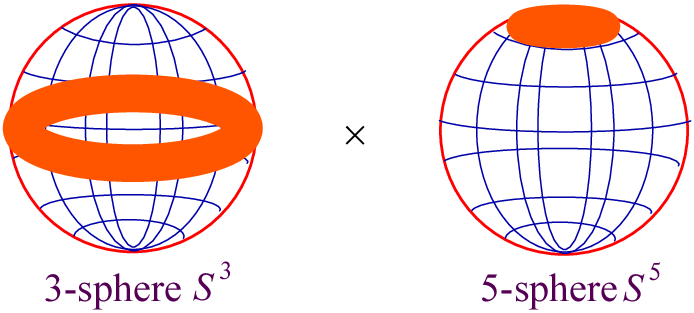}}
\caption{A schematic view of the rotational wave functions of several baryons.
The shaded areas indicate where the wave functions of the eight `Euler' angles that
parameterize the $S^3,S^5$ spheres, are large. Top left: proton, spin up; top right:
$\Omega^-$, spin down-down; bottom left: $\Theta^+$, spin up; bottom, right:
$\Theta^+$, spin up at $N_c=37$.}
\end{figure}

Let us show that the limit $N_c\to\infty$ corresponds to the weak kaon field in the
$\Theta^+$ baryon. To that end we use an alternative parameterization for the meson
field fluctuations about the Skyrmion, suggested by Callan and Klebanov~\cite{CK}:
\beq
U=\sqrt{U_0}\,U_K\sqrt{U_0},
\la{CKpar}\eeq
where $U_K$ is the meson $SU(3)$ unitary matrix which, for small meson fluctuations
$\phi^A$ about the saddle-point Skyrmion field $U_0$ \ur{hedgehog}, is
\bea\la{UK}
&&U_K={\bf 1}_3+i\phi^A\lambda^A, \qquad A=1...8,\\
\n
&&\pi^{\pm}=\frac{\phi^1\pm i\phi^2}{\sqrt{2}},\quad\pi^0=\phi^3,\quad
K^{\pm}=\frac{\phi^4\pm i\phi^5}{\sqrt{2}},\quad
K^0,\overline{K^0}=\frac{\phi^6\pm i\phi^7}{\sqrt{2}},\quad
\eta=\phi^8.
\eea
One can compare \Eq{CKpar} with the rotational Ansatz, $U=RU_0R^\dagger$, and find
the meson fields in baryons corresponding to rotations. In particular,
for rotations ``near the North pole'' {\it i.e.} at small angles $\theta,\phi$,
one finds the kaon field~\cite{DP-05}
\bea
\left.\begin{array}{ccc}K^+&=&-\sqrt{2}\,\sin\frac{P(r)}{2}
\left[\theta\, n_z\!+\!\phi\,(n_x\!-\!in_y)\right]\\
K^0&=&-\sqrt{2}\,\sin\frac{P(r)}{2}
\left[\theta\,(n_x\!+\!in_y)\!-\!\phi\, n_z\right]\end{array}\right\}
=-\sqrt{2}\,\sin\frac{P(r)}{2}\,
(\mbox{\boldmath{$n$}}\cdot\mbox{\boldmath{$\tau$}})
\left(\begin{array}{c}\theta\\ \phi\end{array}\right),
\la{smallK}\eea
meaning that at large $N_c$ the amplitude of the kaon fluctuations in the
prototype ``$\Theta$" is vanishing as $\sim 1/\sqrt{N_c}$. Therefore, at large $N_c$
the rotation is in fact a small-angle precession about the North pole, that
can be studied as a small kaon field fluctuation about the Skyrmion
in a given particular model for the \ECL~\cite{Klebanov,Rho}.
It should be kept in mind, however, that in reality at $N_c\!=\!3$
the rotations by large angles $\theta,\phi$ are not suppressed. It means that
in the real world the kaon field in the $\Theta^+$ is generally not small.

\section{Kaons scattering off the Skyrmion}

As explained in the previous section, at large $N_c$ the kaon field
in the exotic baryon $\Theta^+$ is weak, hence the resonance
should manifest itself in the linear order in the kaon field perturbing
the nucleon which, again at large $N_c$, can be represented by a Skyrmion~\cite{CK,Klebanov}.
In this section we look for the $\Theta^+$ by studying small kaon field
fluctuations about the Skyrmion taking as a model for the \ECL the Skyrme Lagrangian:
\bea\la{action}
S&=&S_{\rm kin}+S_{\rm Sk}+S_{\rm WZ}+S_{m},\\
\la{Skin}
S_{\rm kin}&=&\frac{F_\pi^2}{4}\int\!d^4x\,\Tr L_\mu L_\mu,\qquad L_\mu:=iU^\dagger\partial_\mu U,\\
\la{SSk}
S_{\rm Sk}&=&-\frac{1}{32e^2}\int\!d^4x\,\Tr[L_\mu L_\nu]^2,\\
\la{SWZ}
S_{\rm WZ}&=&\!\frac{N_c}{24\pi^2}\!\!\!\int\!\!d^4x\!\!
\int_0^1 ds\,\epsilon^{\alpha\beta\gamma\delta}
\,\Tr\Bigl(e^{-is\Pi}\partial_\alpha e^{is\Pi}\Bigr)\Bigl(_\beta\Bigr)
\Bigl(_\gamma\Bigr)\Bigl(_\delta\Bigr),\; e^{i\Pi}=U,\\
\la{Sm}
S_m&=&\int\!d^4x\,\frac{m_K^2F_\pi^2}{2}\Tr\left[(U+U^\dagger-2\!\cdot\!{\bf 1}_3)\,
{\rm diag}\left(\frac{m_u}{m_s},\frac{m_d}{m_s},1\right)\right].
\eea
We have written the Wess--Zumino term \ur{SWZ} in the explicit form suggested in
Ref.~\cite{DE}. In the last, symmetry breaking term, we shall put $m_{u,d}=0$.

Following the general approach of Callan and Klebanov~\cite{CK} revived by
Klebanov {\it et al.}~\cite{Klebanov} in the pentaquark era,
we use the parameterization of $U({\bf x},t)$ in the form of \Eq{UK} where
we take the small kaon fluctuation in the form hinted by \Eq{smallK}:
\beq
K^\alpha({\bf x},t)=(\mbox{\boldmath{$n$}}\cdot\mbox{\boldmath{$\tau$}})^\alpha_\beta\,
\zeta^\beta\,\eta(r)\,e^{-i\omega t}
\la{smallK1}\eeq
where $\zeta^\beta$ is a constant spinor. It corresponds to the $p$-wave kaon field.

Expanding the action \ur{action} in the kaon field up to the second order one
obtains~\cite{CK,Klebanov} (we measure $r,t$ in conventional units of
$1/(2F_\pi e)={\cal O}(N_c^0)$)
\bea\la{Sf}
S&=&S_0+S_2,\\
\la{S0}
S_0&=&\frac{2\pi F_\pi}{e}\int\!dr\,r^2\left[\frac{d(r)}{2}\left(1+2s(r)\right)
+s(r)\left(1+\frac{s(r)}{2}\right)\right],\\
\la{S2}
S_2&=&\frac{4\pi F_\pi}{e}\,\zeta^\dagger\zeta \!\int\!dr\,r^2\,\eta(r)
\left\{\omega^2 A(r)- 2\omega\gamma B(r)\right.\\
\n
&&+\left.\left[C(r)\frac{d^2}{dr^2}+D(r)\frac{d}{dr}-V(r)\right]\right\}\eta(r)
\eea
where one introduces short-hand notations:
\bea\n
&& b(r):=P''(r)\sin P(r)+P^{'\,2}(r)\,\cos P(r),\quad c(r):=\sin^2\frac{P(r)}{2},
\quad d(r):=P^{'\,2}(r), \\
\n
&& h(r):=\sin(2P(r))P'(r),\quad s(r):=\frac{\sin^2P(r)}{r^2},\\
\n
&& A(r):=1+2s(r)+d(r),\quad B(r):=-N_ce^2\,\frac{P'(r)\sin^2P(r)}{2\pi^2r^2}, \\
\n
&& C(r):=1+2s(r),\quad D(r):=\frac{2}{r}\left[1+h(r)\right], \\
\n
&& V(r):=-\frac{1}{4}\left[d(r)+2s(r)\right]-2s(r)\left[s(r)+2d(r)\right]
        +2\frac{1+d(r)+s(r)}{r^2}\left[1-c(r)\right]^2\\
&&\qquad\qquad +\frac{6}{r^2}\left[s(r)\left(1-c(r)\right)^2-b(r)(1-c(r))+\half r^2d(r)s(r)\right]
+\mu_K^2.
\la{notations}\eea
Here $\mu_K$ is the dimensionless kaon mass, $\mu_K=m_K/(2F_\pi e)$. The term linear in $\omega$ in \Eq{S2}
arises from the Wess--Zumino term \ur{SWZ}; the function $B(r)$ is the baryons number density in the
Skyrme model. The coefficient $\gamma$ in front of it is unity in the chiral limit but in general is
not universal. In what follows it is useful to analyze the results as one varies $\gamma$ from 0 to 1.

Varying $S_0$ with respect to $P(r)$ one finds the standard Skyrmion profile with $P(0)=\pi$ and
$P(r)\stackrel{r\to\infty}{\longrightarrow}r_0^2/r^2$. Varying $S_2$ with respect to the kaon
field profile $\eta(r)$ one obtains a Schr\"odinger-type equation
\beq
\left\{\omega^2 A(r)- 2\omega\gamma B(r)+\left[C(r)\frac{d^2}{dr^2}+D(r)\frac{d}{dr}-V(r)\right]\right\}\eta(r)=0
\la{Schr}\eeq
where the profile $P(r)$ found from the minimization of $S_0$ has to be substituted.
In the chiral limit ($m_K\to 0$) the equations for $P(r),\,\eta(r)$ are
equivalent to the conservation of the axial current, $\partial_\mu j_{\mu\,5}^A=0$
since it is the equation of motion for the Skyrme model.

If $m_K=0$, the $SU(3)$ symmetry is exact, and a small and slow rotation in the strange direction
must be a zero mode of \Eq{Schr}. Indeed, one can easily check that
\beq
\eta_{\rm rot}(r)=\sin\frac{P(r)}{2}
\la{eta-rot}\eeq
is a zero mode of the square brackets in \Eq{Schr} and hence a zero mode of the full
equation with $\omega\!=\!0$. If in addition the Wess--Zumino coefficient $\gamma$ is set
to zero, this mode is twice degenerate. These states are the large-$\!N_c$ prototypes of
$\Lambda$ (strangeness $S\!=\!-1$) and $\Theta^+$ ($S\!=\!+1$)~\cite{CK,Klebanov}.
At $\gamma >0$ the two states split: $\Lambda$ remains a pole of the scattering amplitude
at $\omega\!=\!0$, and $\Theta^+$ moves into the lower semi-plane of the complex $\omega$ plane.
If $m_K>0$ the pole corresponding to the $\Lambda$ moves to $\omega\!<\!0$ remaining on
the real axis, whereas the $\Theta^+$ pole remains in the lower semi-plane with
${\rm Re}\;\omega\!>0$ and ${\rm Im}\;\omega\!<0$. Both poles are singularities of the
same analytical function {\it i.e.} the scattering amplitude, see below. It is amusing
that $\Lambda$ ``knows'' about $\Theta^+$ and its width through analyticity.

In what follows we shall carefully study the solutions of \Eq{Schr} and in particular the
trajectory of the $\Theta^+$ pole, by combining numerical and analytical calculations.
In numerics, we use the conventional choice of the constants in the Skyrme model: $F_\pi=64.5$ MeV
({\it vs} 93 MeV experimentally) and $e=5.45$. These values fit the nucleon mass $m_N=940\,{\rm MeV}$
(with the account for its rotational energy) and the mass splitting between the nucleon and
the $\Delta$-resonance~\cite{AdkinsNappiWitten}. These were the values used also by Klebanov
{\it et al.}~\cite{Klebanov} who solved numerically \Eq{Schr} and found the phase shifts
$\delta(\omega)$ defined from the large-$r$ asymptotics of the solutions of \Eq{Schr} regular
at the origin,
\beq
\eta_{\rm as}(r)=\frac{k r+i}{r^2}e^{i k r+i\delta(\omega)}
+\frac{k r-i}{r^2}e^{-ik r-i\delta(\omega)},\qquad (k=\sqrt{\omega^2-m_K^2}),
\la{asympt}\eeq
being a superposition of the incoming and outgoing spherical waves.
%Corrections to \Eq{asympt} are ${\cal O}(1/r^4)$ related to the decrease
%of the profile function $P(r)\sim r_0^2/r^2$.
At $\gamma=1$ and physical $m_K=495\,{\rm MeV}$, the phase shift $\delta(\omega)$
has been found in Ref.~\cite{Klebanov} to be less than $45^{\small o}$
in the range of interest. This have lead the authors to the conclusion that
$\Theta^+$ does not exist in the Skyrme model, at least in the large $N_c$ limit and small $m_K$.
We reproduce their phase shifts with a high accuracy (as well as the phase shifts
studied in Ref.~\cite{WW} for another choice of the Skyrme model parameters) but
come to the opposite conclusions.

%Quotation from Klebanov et al. Abstract:
%However, for small mK, we find no S=+1 kaon bound states or resonances in the spectrum, confirming previous work.
%This suggests that, at least for large N and small mK, the exotic state may be an artifact of the rigid rotator
%approach to the Skyrme model. An S=+1 near-threshold state comes into existence only
%for sufficiently large SU(3) breaking. If such a state exists, then it has the expected quantum
%numbers of $\Theta+$: I=0, Image  and positive parity.

In a situation when there is a resonance and a potential scattering together,
the phase shift does not need to go through $90^{\rm\small o}$ as it would be requested
by the Breit--Wigner formula for an isolated resonance. A far better and precise way
to determine whether there is a resonance, is to look not into the phase shifts
but into the singularities of the scattering amplitude in the complex energy plane.
A resonance is, by definition, a pole of the scattering amplitude in the lower semi-plane
on the second Riemann sheet:
\beq
\sqrt{s}_{\rm pole} = m_{\rm res} - i\frac{\Gamma}{2}
\la{pole}\eeq
where $m_{\rm res}$ is the resonance mass and $\Gamma$ is its width.

The scattering amplitude $f(\omega)$ and the scattering matrix $S(\omega)$
(which in this case has only one element) are defined as
\beq
f(\omega)=\frac{1}{2i k}\left(e^{2i\delta(\omega)}-1\right),
\qquad S(\omega)=e^{2i\delta(\omega)}.
\la{S-matrix}\eeq
A standard representation for the scattering amplitude is
\beq
f=\frac{1}{g(\omega)-ik}, \qquad g(\omega)=k\cot\delta(\omega).
\la{g-function}\eeq
This representation solves the unitarity condition for the $S$-matrix: $g$ is real
on the real $\omega$ axis. The function $g(\omega)$ does not have cuts related
to the $KN$ thresholds and $\omega^{2l}g(\omega)$ is Taylor-expandable at small $\omega$,
therefore it is a useful concept~\cite{LL}.

For $\omega$ in the lower complex semi-plane the first term in \Eq{asympt} becomes
a rising exponent of $r$, and the second term becomes a falling exponent. Since the
$S$ matrix is proportional to the ratio of the coefficient in front of
$\exp(-ikr)$ to that in front of $\exp(ikr)$, the pole of the $S$ matrix and hence
of the scattering amplitude corresponds to the situation where the wave function
$\eta(r)$ regular at the origin, has no falling exponent at $r\to\infty$ but only
a rising one. Physically, it corresponds to a resonance decay producing outgoing waves only.

For the conventional choice of the parameters we find the $\Theta^+$ pole position at
\beq
\sqrt{s}_{\rm pole} = \left\{\begin{array}{ccc}(1115 - 145 i)\; {\rm MeV} & {\rm for}\; m_K=0 &
({\rm threshold\;at}\; 940\;{\rm MeV})\\
(1449 -44 i)\;{\rm MeV} & {\rm for}\; m_K=495\;{\rm MeV} &
({\rm threshold\;at}\; 1435\;{\rm MeV})\end{array}\right.
\la{pole-position}\eeq
We have recalculated here the pole position in $\omega$ to the relativistic-invariant
$KN$ energy $s=m_N^2+2m_N\omega+m_K^2$. It is a perfectly normal resonance in the
strong interactions standards with a width of 90 MeV. It would be by all means seen
in a partial wave analysis (see Fig.~9, left) or just in the $KN$ total $T\!=\!0$ cross section
which we calculate from the well-known equation $\sigma=4\pi(2j+1)|f|^2
=\frac{4\pi}{k^2}(2j+1)\sin^2\delta$ (Fig.~9, right).

%%%%%%%%%%%%%%%
%% FIGURE 9  %%
%%%%%%%%%%%%%%%
\begin{figure}[]
\centerline{\hbox{\epsfig{figure=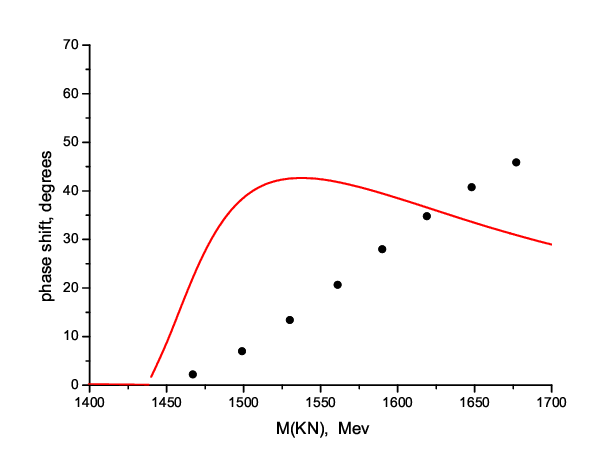,width=6.9cm}}\hspace{-0.5cm}
\hbox{\epsfig{figure=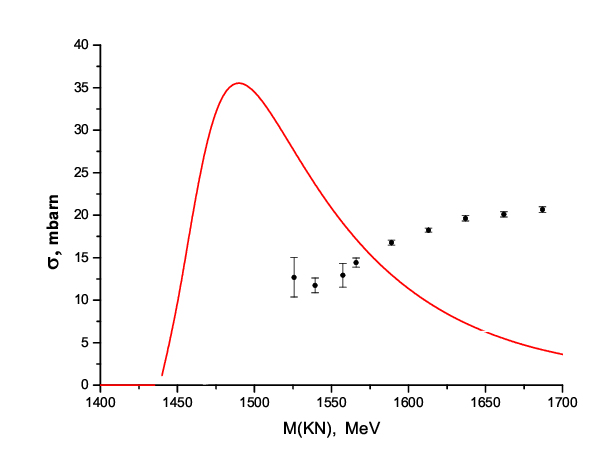,width=6.9cm}}}
\caption{{\it Left}: the $T\!=\!0,L\!=\!1$ $KN$ scattering phase as function of the $KN$
invariant mass in the Skyrme model in the large-$\!N_c$ limit (it coincides with
the phase found in Ref.~\cite{Klebanov}), compared to the result of
the partial wave analysis~\cite{Arndt} shown by dots. {\it Right}: the ensuing $KN$
cross section in this partial wave exhibits a strong resonance around 1500 MeV,
whereas the experimental data~\cite{Dover} for the sum over all partial waves
shows no signs of a resonance.}
\end{figure}
At the maximum the cross section is as large as 35 mb, and it is
a contribution of only one particular partial wave $P_{01}$! Needless to say, such a strong
resonance is not observed. Varying the parameters of the Skyrme model~\cite{Klebanov,WW}
or modifying it~\cite{Rho} can make the exotic resonance narrower or broader
but one cannot get rid of it. The reason is very general: poles in the scattering amplitude
do not disappear as one varies the parameters but move in the complex plane.

One can check it in a very precise way by, say, varying artificially the coefficient
in front of the Wess--Zumino term $\gamma$ from 0 to 1. At $\gamma=0$ there is certainly
an exotic bound state at $\omega=0$ corresponding to the rotational zero mode \ur{eta-rot}.
At $\gamma>0$ the position of the pole of the $KN$ scattering amplitude moves into the
complex $\omega$ plane such that
\bea\n
{\rm Re}\;\omega_{\rm pole} &=& a_1 \gamma + a_3 \gamma^3 +\ldots,\\
{\rm Im}\;\omega_{\rm pole} &=& b_2 \gamma^2 + b_4 \gamma^4 +\ldots
\la{gamma-expansion}\eea
with analytically calculable coefficients in this Taylor expansion (we give
explicitly the leading coefficients in Section 7). By comparing the
numerical determination of the pole position with the analytical expressions we
trace that the $\Theta^+$ pole \ur{pole-position} is a continuous deformation
of the rotational mode, see Fig.~10.\\
\vskip -0.4true cm

Thus, the prediction of the Skyrme model is not that there is no $\Theta^+$
but just the opposite: {\bf there must be a very strong resonance}, at least
when the number of colours is taken to infinity. Since this prediction is
of general nature and does not rely on the specifics of the Skyrme model,
one must be worried why a strong exotic resonance is not observed experimentally!

%%%%%%%%%%%%%%%
%% FIGURE 10 %%
%%%%%%%%%%%%%%%
\begin{figure}[t]
\centerline{\hbox{\epsfig{figure=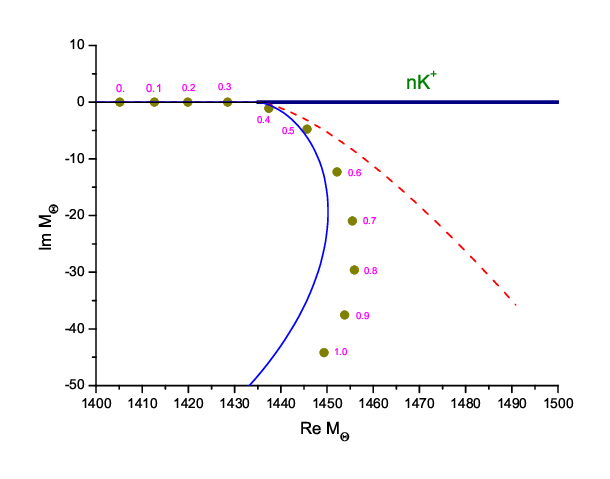,width=8cm,height=6cm}}}
\caption{Trajectory of the pole in the $KN$ scattering amplitude for ``realistic'' parameters
of the Skyrme model and physical $m_K=495\,{\rm MeV}$ at $\gamma=0,\,0.1,\ldots,1.0$. The dashed
and solid lines show the analytical calculation of the pole position in the first and second
orders in $\gamma$, respectively.}
%\vskip -0.4true cm
\end{figure}

The answer is that the large-$N_c$ logic in general and the concrete Skyrme model
in particular grossly overestimate the resonance width (we explain it in
the next Sections). The resonance cannot disappear but in reality it becomes
very narrow, and that is why it is so difficult to observe it.

One may object that the Skyrme model is a model anyway, and a modification of its
parameters or a replacement by another chiral model can lead to an even larger
width, say, of 600 MeV instead of 90 MeV obtained here from the ``classical'' Skyrme model.
However, as we argue in Section 9, going from $N_c\!=\!\infty$ to the real world
at $N_c\!=\!3$ reduces the width by at least a factor of 5. Therefore, even a 600-MeV
resonance at $N_c\!=\!\infty$ would become a normal 120-MeV resonance in the real world
and would be observable.

Thus, the only way how a theoretically unavoidable resonance can escape observation
is to become very narrow. We remark that the reanalysis of the old $KN$ scattering
data~\cite{Arndt-03} shows that there is room for the exotic resonance with a mass
around 1530 MeV and width below 1 MeV.

\section{Physics of the narrow $\Theta^+$ width}

Quantum field theory says that baryons cannot be $3Q$ states only but necessarily
have higher Fock components due to additional $Q\bar Q$ pairs; it is only a quantitative
question how large are the $5Q,\;7Q,...$ components in ordinary baryons. Various
baryon observables have varying sensitivity to the presence of higher Fock
components. For example, the fraction of the nucleon momentum carried by antiquarks
is, at low virtuality, less than 10\%. However, the nucleon $\sigma$-term
or nucleon spin are in fact dominated by antiquarks~\cite{DPP-89,WY}.
Both facts are in accord with a normalization of the $5Q$ component of the nucleon
at the level of 30\% from the $3Q$ component, meaning that 30\% of the time nucleon
is a pentaquark!

As to the exotic $\Theta^+$ and other members of the antidecuplet, their
{\em lowest} Fock component is the $5Q$ one, nothing terrible. However it has
dramatic consequences for the antidecuplet decay widths.

To evaluate the width of the $\Theta^+\to K^+n$ decay one has to compute the
transition matrix element of the strange axial current,
$<\!\Theta^+|\bar s\gamma_\mu\gamma_5u|n\!>$. There are two contributions
to this matrix element: the ``fall apart'' process (Fig.~10, A) and the ``5-to-5''
process where $\Theta^+$ decays into the $5Q$ component of the nucleon (Fig.~10, B).
One does not exist without the other: if there is a ``fall apart'' process it means that there
is a non-zero coupling of quarks to pseudoscalar (and other) mesons, meaning
that there is a transition term in the Hamiltonian between $3Q$ and $5Q$ states
(Fig.~10, C). Hence the eigenstates of the Hamiltonian must be a mixture of
$3Q,5Q,...$ Fock components. Therefore, assuming there is process A, we have to
admit that there is process B as well. Moreover, each of the amplitudes A and B
are not Lorentz-invariant, only their sum is. Evaluating the ``fall-apart'' amplitude
and forgetting about the ``5-to-5'' one makes no sense.

%%%%%%%%%%%%%%%
%% FIGURE 11 %%
%%%%%%%%%%%%%%%
\begin{figure}[]
\centerline{\hbox{\epsfig{figure=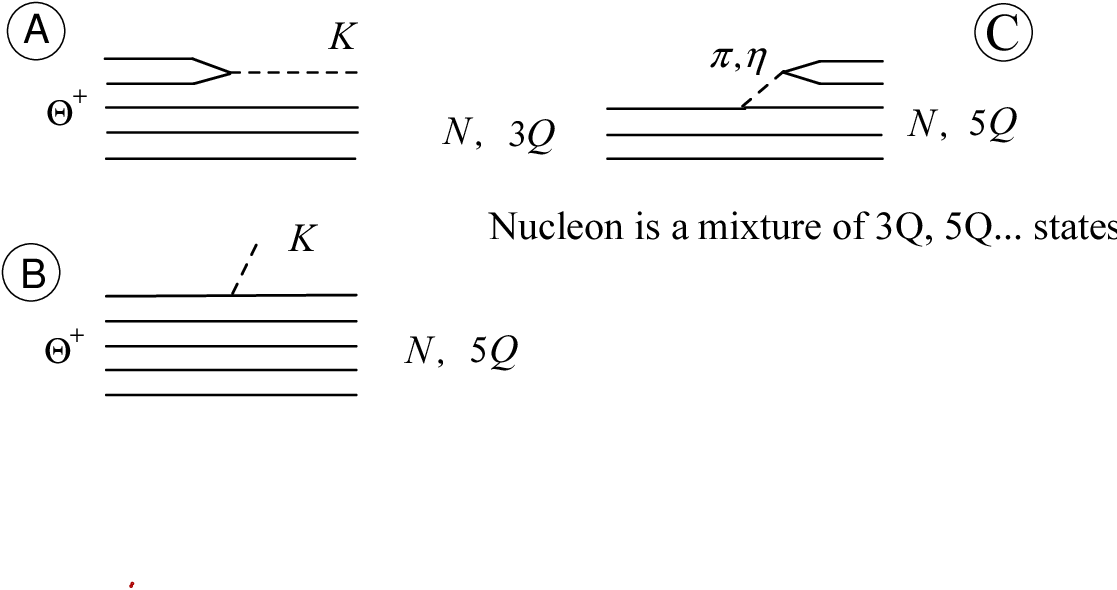,width=10cm}}}
\vskip -1.7true cm

\caption{``Fall-apart'' (A) and ``5-to-5'' (B) contributions to the $\Theta^+\to K^+n$ decay.}
\end{figure}

A convenient way to evaluate the sum of two graphs, A and B, in the chiral limit
is to go to the infinite momentum frame (IMF) where only the process B
survives, as axial (and vector) currents with a finite momentum transfer do not
create or annihilate quarks with infinite momenta. In the IMF the fall-apart
process A is exactly zero in the chiral limit. The baryon matrix elements
are thus non-zero only between Fock components with {\em equal number} of
quarks and antiquarks.

The decays of ordinary (nonexotic) baryons are mainly due to the $3Q\to 3Q$
transitions with a small (30\%) correction from $5Q\to 5Q$ transitions and even smaller
corrections from higher Fock components, just because the $3Q$ components dominate.
The nucleon axial constant is thus expected to be of the order of unity and indeed
$g_A(N)=1.27$.

However the $\Theta^+$ decay is dominated by the $5Q\to 5Q$ transition,
and {\bf the transition axial constant is suppressed to the extent the $5Q$
component in the nucleon is suppressed with respect to the $3Q$ one}~\cite{DP-03}.

A quantitative estimate of this effect can be made in a relativistic model
(since pair creation and annihilation is a relativistic effect) where it is
possible to calculate both the $3Q$ and the $5Q$ wave functions of the nucleon and
the $5Q$ wave function of the $\Theta^+$. We claim that in any such model
of baryons the $\Theta^+$ will be narrow if the model tells that the $5Q$ component
in the nucleon is suppressed with respect to the $3Q$ one. An example is provided by
the Chiral Quark Soliton Model where, indeed, the $5Q\to 5Q$ axial constant has
been estimated as $g_A(\Theta\to KN)\approx 0.14\;$~\cite{DP-05,Lorce} yielding
\beq
\Gamma_\Theta \approx 2\;{\rm MeV}.
\la{GaTh}\eeq
Apart from the suppression of general nature discussed above there is an additional
suppression of $g_A(\Theta\to KN)$ due to $SU(3)$ group factors in the $\overline{\bf 10}
\to {\bf 8}$ transition. This estimate has been performed assuming the chiral limit
($m_K=0$) and zero momentum transfer in the decay. In more realistic settings the width
can only go down. A recent calculation with account for the decay via higher Fock components
and also for $m_s$ corrections gives $\Gamma_\Theta=0.7\,{\rm MeV}\;$~\cite{LKG}.

As stressed in Ref.~\cite{DPP-97} where the narrow $\Theta^+$ has been first predicted,
in the imaginary nonrelativistic limit when ordinary baryons are made of three quarks
only with no admixture of $Q\bar Q$ pairs the $\Theta^+$ width tends to zero strictly. \\

It may seem that by the same argument all members of the exotic multiplets $\adec$, $\pletthree$ {\it etc.},
should be very narrow resonances but the above argument applies only to the transitions with
an emission of one pseudoscalar meson. As a matter of fact it applies also to the $BBV$
transitions where $V$ is a vector meson that couples to baryons via the
conserved vector current; such transitions are also expected to be strongly
suppressed~\cite{PolyakovRathke}, and the calculations~\cite{LKG} support it. However, the
argument does not work for transitions with two or more pseudoscalar mesons emission.
Therefore, if the phase volume allows for a decay of an exotic resonance to two or
more mesons, the width does not need to be particularly narrow; it should be studied
on case-to-case basis.\\

We now return to the Skyrme model and discuss why does it give a wide $\Theta^+$
in the large $N_c$ limit. As explained in Section 1, the Skyrme model is an idealization
of nature: It implies that the chiral field is broad, the valence quarks are close
to the negative-energy Dirac sea and cannot be separated from the sea, whereas the distortion
of the sea is large. The number of $Q\bar Q$ pairs originating from the strongly
deformed sea is ${\cal O}(1)$, times $N_c$. For example, it has been known for quite
a while~\cite{BEK} that the fraction of nucleon spin carried by valence quarks is zero
in the Skyrme model. Therefore, the Skyrme model implies the situation which is the
opposite extreme from the nonrelativistic quarks where there are $N_c$ valence quarks
and zero $Q\bar Q$ pairs. From the viewpoint of parton distributions, both limits
are discussed in some detail in Ref.~\cite{DPPPW}.

Therefore, the Skyrme model fails to accommodate the real-world physics explaining
the narrow $\Theta^+$, in two essential points:
\begin{itemize}
\item At large $N_c$ justifying the study of the $\Theta^+$ resonance from the kaon-Skyrmion
scattering both the nucleon and the $\Theta^+$ have an equal (and infinite) number of
$Q \bar Q$ pairs; hence the $\Theta\to KN$ transition is not suppressed at all.
[This is why we have obtained a large 90 MeV width in Section 5]\\

\item Even if one takes a moderate $N_c\!=\!3$, the Skyrme model implies that the
$5Q$ component of the nucleon remains large, and there is no argument why the
$\Theta^+$ width should be particularly small, although it must be less than in
the infinite-$\!N_c$ limit.
\end{itemize}

Having this understanding in mind, in the next section we return to the Skyrme model
to see if it is possible to play with its parameters in such a way that it would
mimic to some extent the nonrelativistic limit. Then the ``$\Theta^+$'' of such a model
should be narrow even if $N_c$ is large.

\section{Getting a narrow $\Theta^+$ in the Skyrme model}

Unfortunately, there are very few handles in the Skyrme model to play with. In fact,
there are only three constants: $F_\pi$, $e$ and the coefficient in front of the Wess--Zumino term,
$\gamma$. The last one is unity in the ideal case but is not universal if, for example, $m_K\neq 0$.
A general statement is that $\gamma$ decreases as $m_K$ increases. The constant $F_\pi$ has
to be taken 1.5 times less than its physical value to fit 940 MeV for the nucleon mass.
The dimensionless coefficient $e$ is also rather arbitrary, it is not being fixed from
the experimental $d$-wave pion scattering. Therefore, we feel free to modify these constants
at will, in order to make a theoretical point. The models we are going to present are not
realistic, of course. To get more realistic models, one has either to add vector mesons~\cite{Rho}
or take the Chiral Quark Soliton Model, or do something else.

\subsection{Vanishing $m_\Theta-m_N$, vanishing $\Gamma_\Theta$}

We start with a simple exercise, making $\gamma$ a small number. To simplify
the presentation we take the chiral limit $m_K=0$ but give the final results for
$m_K\neq 0$ at the end of this subsection.

At small $\gamma$, one can compute everything analytically. In particular, one can find
the regular function $g(\omega)$ \ur{g-function} in the range of interest $\omega=k\sim \gamma$.
This is done by comparing the asymptotics of the wave function \ur{asympt} in the range
$kr\ll 1$ but $r\gg r_0$ where $r_0$ is the coefficient in the asymptotics of the profile function
$P(r)\approx r_0^2/r^2$, with the asymptotics of $\eta(r)$ at $r\gg r_0$ being
\beq
\eta(r)=\frac{C_1}{r^2}+C_2\,r,\quad {\rm from\;where}\;\; g(\omega)=\frac{3C_2}{C_1\omega^2}.
\la{g1st}\eeq
The coefficients $C_{1,2}$ are found from the following considerations. In the range of
interest, $k\sim \gamma \ll 1$, the wave function $\eta(r)$ differs from the
rotational wave function $\eta_{\rm rot}(r)$ \ur{eta-rot}, being the exact solution
of \Eq{Schr} at $\omega=0$, by terms of the order of $\gamma^2$. $\eta_{\rm rot}(r)$ falls
off at large distances as $r_0^2/2r^2+0\cdot r$. Therefore,
$C_1=r_0^2/2+{\cal O}(\gamma^2)$ whereas $C_2={\cal O}(\gamma^2)$ and can be computed, in
the leading order, as a matrix element \ur{S2} with $\eta(r)$ substituted by $\eta_{\rm rot}(r)$.
We find
\beq
C_1=\frac{r_0^2}{2}+{\cal O}(\gamma^2),\qquad
C_2=\frac{1}{6\pi r_0^2}\left(\gamma e^2 N_c\,\omega - \tilde I_2\,\omega^2\right) = {\cal O}(\gamma^2)
\la{C12}\eeq
where the second moment of inertia $\tilde I_2$ arises here as
$$
\tilde I_2=4\pi\int\!drr^2\eta_{\rm rot}A(r)\eta_{\rm rot}(r).
$$
The physical moment of inertia (in MeV) is obtained from the dimensionless
$\tilde I_2$ as $I_2 = \tilde I_2/(8e^3F_\pi)$.

Technically, we obtain $C_2$ by the following trick: We integrate \Eq{Schr}
multiplied to the left by $\eta_{\rm rot}(r)$ from zero to some $r\gg r_0$,
and subtract the same integral with $\eta(r)$ and $\eta_{\rm rot}(r)$ interchanged.
The terms with the first and second derivatives of $\eta,\eta_{\rm rot}$ become
a full derivative that can be evaluated at the integration end point, while
terms with no derivatives are fast convergent such that one can extend the
integration range to infinity and also replace $\eta\to\eta_{\rm rot}$ in the
leading order.

The poles of the scattering amplitude are found from the equation $g(\omega)=i\omega$.
Using \Eqs{g1st}{C12} we get the real and imaginary parts of the pole position:
\bea\n
\Delta &=&\gamma\omega_{\rm rot}={\cal O}(\gamma),\qquad
\omega_{\rm rot}=\frac{N_c}{4I_2},\\
%\frac{2F_\pi e^3 N_c}{\tilde I_2}
\Gamma &=& \frac{8\pi F_\pi^2r_0^4}{\gamma N_c}\,\Delta^3={\cal O}(\gamma^2) \ll \Delta,
\la{DG}\eea
where $\Delta=m_{\Theta}-m_N$, $\Gamma$ and $r_0$ are in physical units.
These are actually the first terms in the expansion \ur{gamma-expansion}.
We have developed a perturbation theory in $\gamma$ and found analytically
higher order terms as well. The first few orders follow the numerical
determination of the pole position all the way up to $\gamma\!=\!1$, see
Fig.~10. It makes us confident that even at finite width the $\Theta^+$ resonance
is a continuous deformation of the rotational would-be zero mode.

Since in the limit of small $\gamma$ everything is analytically calculable,
one can check various facts. In particular, we have computed the transition axial constant
$g_A(\Theta\to KN)$ from the asymptotics \ur{asympt} of the kaon scattering wave $\eta(r)$.
It gives the needed (massless) kaon pole $1/(\omega^2-k^2)$; the axial constant is
the residue of this pole, for more details see Section 8. The overall spinor coefficient
$\zeta^\alpha$ in the kaon wave is fixed from the quantization condition requesting
that a state with strangeness $S=+1$ and exoticness $X=1$ is formed~\cite{CK}. It gives
\beq
\langle \zeta^\dagger_\alpha\zeta^\alpha\rangle=\frac{8}{\gamma N_c}.
\la{normzeta}\eeq
We obtain
$$
%\left.
g_A(\Theta\to KN)
%\right|_{\gamma\to 0}
=8\pi \frac{F_\pi^2r_0^2}{\sqrt{2\gamma N_c}}.
$$
Given the axial constant, the pseudoscalar coupling $G(\Theta\to KN)$ can be found
from the Goldberger--Treiman relation
$$
G(\Theta\to KN)=\frac{m_N\,g_A(\Theta\to KN)}{F_\pi}.
$$
The pseudoscalar coupling and the decay phase volume determines the $\Theta^+$ width:
$$
\Gamma=\frac{G^2(\Theta\to KN)\Delta^3}{4\pi m_N^2}
=\frac{8\pi F_\pi^2r_0^4}{\gamma N_c}\,\Delta^3.
$$
Comparing it with the determination of $\Gamma$ from the pole position \ur{DG}, we see that
the two ways of determining the width coincide!\\

We can determine the width in a third way -- from the {\em radiation} of the
kaon field by the resonance~\cite{DP-Regge}. According to Bohr's correspondence principle,
the quantum width is determined as the inverse time during which a resonance looses, through
classical radiation, the energy difference between the neighbour states:
\beq
\Gamma = \frac{W}{\Delta}
\la{Bohr}\eeq
where $W$ is the radiation intensity, {\it i.e.} the energy loss per unit time.
Strictly speaking, Bohr's principle is semiclassical and applies only to the
decays of the highly excited levels. In our case, however, we linearize in the kaon
field at large $N_c$, therefore it is essentially a problem for a set of harmonic
oscillators for which semiclassics is exact starting from the first excited level,
here the $\Theta^+$.

To find the radiation intensity $W$, we look for the solution of \Eq{Schr} with only
the outgoing wave in the asymptotics, $\eta(r)=c(kr+i)e^{ikr}/r^2$. The coefficient
$c$ is found from matching it in the range $r_0\ll r \ll 1/k$ with the solution regular
at the origin. In the leading order in $\gamma$ it is the rotational mode
$\eta_{\rm rot}(r)\approx r_0^2/2r^2$; it gives $c=-ir_0^2/2$. The radiation intensity
corresponding to this outgoing wave is found as the flux through a distant surface
of the Pointing vector
$$
T_{0i}\quad\stackrel{r\to\infty}{=}\quad\frac{F_\pi^2}{2}\left(\partial_0K^\dagger_\alpha \partial_i K^\alpha
+\partial_i K^\dagger_\alpha \partial_0K^\alpha\right)\quad\stackrel{r\to\infty}{=}\quad
\frac{F_\pi^2r_0^4k^4}{4r^2}\zeta^\dagger_\alpha \zeta^\alpha.
$$
Using the normalization \ur{normzeta} we find the radiation intensity
$$
W=\int d^2S_i\,T_{0i}=4\pi\frac{F_\pi^2r_0^4k^4}{4}\frac{8}{\gamma N_c}
$$
where in the massless kaon limit $k=\omega=\Delta$. Note that the $k^4$ dependence
is typical for the dipole radiation; it is dipole as we look for kaon radiation in $p$-wave.
[Another characteristic feature of the dipole radiation -- the $\cos^2\theta$ angular
dependance -- is not seen here because we have in fact averaged over the $\Theta^+$ spin.]
From Bohr's equation \ur{Bohr} we obtain the $\Theta^+$ width
$$
\Gamma=\frac{W}{\Delta}=\frac{8\pi F_\pi^2r_0^4}{N_c\gamma}\,\Delta^3
$$
again coinciding with the determination of the width from the pole position,
\Eq{DG}. One can also compute the width as the inverse time during which
one unit of strangeness is lost through kaon radiation, with the same result.
Yet another (a 5th!) way of computing $\Gamma$ -- from the asymptotics of the classical
profile function of the Skyrmion -- will be presented in Section 9.

The derivation of $\Gamma$ and $\Delta$ can be repeated for $m_K\neq 0$ in which case we find
\beq
\Delta =\Delta_0\left(\half+\sqrt{\frac{1}{4}+\frac{b\,m_K^2}{\Delta_0^2}}\right),\qquad
\Gamma =\Gamma_0\,\frac{\Delta_0}{2\Delta-\Delta_0}\frac{(\Delta^2-m_K^2)^{\frac{3}{2}}}{\Delta_0^3},
%\qquad G = G_0\sqrt{\frac{\Delta_0}{2\Delta-\Delta_0}},
\la{DGmK}\eeq
where $\Delta=m_\Theta-m_N$ when $m_K\neq 0$ while the subscript 0 refers to the case of $m_K=0$.
It is remarkable that the imaginary part of the pole position $\Gamma$ apparently ``knows''
-- through unitarity -- about the decay phase volume clearly visible in \Eq{DGmK}.
The numerical coefficient $b$ is defined as
$b=4\pi\int\!drr^2\eta_{\rm rot}^2(r)/\tilde I_2\approx 0.705.$
\vskip 0.3true cm

To conclude this subsection: in the case when the $\Theta^+$ width is made small,
we have determined it in three independent ways: {\it i}) from the pole position
in the complex energy plane, {\it ii}) from the axial constant and by using the
Goldberger--Treiman relation, {\it iii}) from the semiclassical radiation theory.
All three calculations lead to the same expression for the width $\Gamma$.

\subsection{Finite $m_\Theta-m_N$, vanishing $\Gamma_\Theta$}

The analytical equations of the previous subsection remain accurate as long as the imaginary
part of the pole position is much less than the real part, that is insofar as
$\Gamma\ll\Delta$. From \Eq{DG} one infers that actually this condition is $\gamma e^2N_c\ll 1$
where $e^2$ is the inverse coefficient in front of the Skyrme term in the action,
and $\gamma$ is the coefficient of the Wess--Zumino term. If one likes to
fix once and forever $\gamma=1$ (say, from topology arguments), one is still able
to support the regime $\Gamma\to 0$ but $\Delta={\rm const}$ by rescaling the
other two constants of the Skyrme model. Namely, we consider the following regime:
\beq
F_\pi = F_0 \beta^{-3},\quad e=e_0 \beta,\qquad \beta\to 0, \qquad \gamma = \gamma_0=1.
\la{beta1}\eeq
The Skyrmion mass, its size and moments of inertia scale then as
$$
m_N\sim\frac{F_\pi}{e}\propto\frac{1}{\beta^4}\to \infty,\qquad r_0\sim\frac{1}{F_\pi e}\propto \beta^2\to 0,
\qquad I_{1,2}\sim\frac{1}{F_\pi e^3}\propto{\rm const.},
$$
such that the splittings between rotational $SU(3)$ multiplets remain fixed,
$m_\Delta-m_N\sim {\rm const.}$, and
\beq
\Delta=m_{\Theta}-m_N=\frac{N_c}{4I_2}\sim N_cF_\pi e^3\propto{\rm const.},
\qquad {\rm but}\;\;
\Gamma_\Theta \sim N_c^2F_\pi e^5\propto \beta^2\to 0\,.
\la{scaling}\eeq
We stress that $\Theta^+$ becomes stable in this regime not because the decay
phase volume tends to zero (which would have been trivial but it is not the case here)
but because the Skyrmion size $r_0$ is small. Taking $r_0$ to zero we mimic to some
extent the limit of nonrelativistic quarks in the Skyrme model, where we expect a narrow width.
Since the Skyrme model is opposite in spirit to the nonrelativistic quarks (see Sections
1 and 6) it is difficult to achieve this limit. Indeed, the regime \ur{beta1} is not
too realistic. However it serves well to illustrate the point: When $\Gamma$ is small,
the real part of the pole position coincides with the rotational frequency
\beq
m_{\Theta}-m_N =\omega_{\rm rot}=\frac{N_c}{4I_2}
\la{rot-split}\eeq
as it follows from the quantization of the $SU(3)$ rotations, \Eq{antiten_oct}.

Is it a coincidence? Probably, not: $\Theta^+$ {\em is} an $SU(3)$ rotational
excitation of the nucleon. (At large $N_c$ the rotation is more like a precession
near the ``North pole'' but nevertheless.) It remains a (deformed) rotational state
even in the worse case scenario provided by the Skyrme model where at ``realistic''
parameters it becomes a broad and hence strong resonance but then what should be
called the resonance mass becomes ambiguous. Its precise determination is then from
the pole position which is away from the real axis, such that the real part of the
pole position does not need to coincide with the rotational splitting just for the
trivial reason that the imaginary part is large.

Therefore, the key issue is the resonance width. On the one hand, a rotating body
must radiate, in this case the kaon field. Since in the $\left.\frac{1}{2}\right.^+
\to\left.\frac{1}{2}\right.^+$ transition the kaon is in the $p$ wave, the $\Theta^+$ width
is entirely due to dipole radiation. The dipole radiation intensity is proportional to
$(\ddot d)^2=\omega^4 d^2$ at small frequencies where $d$ is the dipole moment.
Generically, $d\sim r_0$ where $r_0$ is the characteristic size of the system.
The Skyrme model illustrates the generic case, therefore the only way to suppress
the dipole radiation at fixed $\omega=\Delta$ is to shrink the size $r_0$ as we have
done above. On the other hand, in our case it is the {\em transition} dipole moment
$d$ corresponding to strangeness emission, which can be, in principle, much less than $r_0$,
even zero. We have argued in Section 6 that in the real world the transition dipole moment
is small as the nucleon is essentially nonrelativistic and hence has a small $5Q$ component.
If the width is small, we see no reasons why would not the real part of the $\Theta^+$ pole
coincide with the rotational splitting of the $SU(3)$ multiplets~\footnote{A quantum-mechanical
counter-example by T. Cohen~\cite{Cohen-counter} does not seem to capture the necessary physics
as the spectrum there is discrete and there is no radiation.}.

\section{Goldberger--Treiman relation and the $\Theta^+$ width}

In this section we show that the $\Theta^+$ width can be expressed through the
transition axial constant {\em provided} the width is small. We reaffirm the
validity of a modified Goldberger--Treiman relation between the axial and pseudoscalar
$\Theta KN$ constants in the chiral limit. We derive these relations in the
framework of the Skyrme model where all equations are explicit. However, these
relations are, of course, of a general nature.

If $\Theta^+$ is a narrow and well-defined state one can define the transitional axial
$g_A=g_A(\Theta\!\to\!KN)$ and pseudoscalar $G=G(\Theta\!\to\!KN)$ constants as
\beq
\frac{\varepsilon_{\alpha\beta}}{\sqrt{2}}\langle \Theta^+_k|j^\alpha_{\mu 5}(x)|N^{\beta,i}\rangle=
e^{i q\cdot x}
\bar{u}(\Theta,k)\left(\frac{\gamma_\mu}{2}g_A-\frac{q_\mu}{q^2}F_\pi G\right)\,\gamma_5\,u(N,i)
\la{gandG}\eeq
where $i,k=1,2$ are the nucleon and $\Theta^+$ spin projections, $\alpha,\beta=1,2$ are the
isospin projections of the nucleon and of the kaon current; we are interested in the isospin $T\!=\!0$
channel. Finally, $u,\bar u$ are $N$ and $\Theta^+$ 4-spinors.
We assume that they obey the non-relativistic normalization $\bar{u}(i)u(k)=\delta_{ik}$.
In the non-relativistic limit appropriate at large $N_c$ one has:
\beq
\bar{u}(k)\gamma_5 u(i)=\psi^*(k)\frac{{\bf q}\cdot{\bf \sigma}}{2m}\psi(i)
=\frac{1}{2m}({\bf q}\cdot{\bf \sigma})^i_k
\la{nonrel}\eeq
where $\psi(i)$ is a non-relativistic 2-spinor with polarization $i$.

The modified Goldberger--Treiman relation follows immediately from the conservation of
the axial current, $\partial_\mu j^\alpha_{\mu 5}=0$:
$$
g_A(\Theta\!\to\! KN)(m_N+m_\Theta)=2G(\Theta\!\to\! KN)F_{\pi}.
$$
It should be stressed that it holds true even if $m_\Theta$ differs significantly from $m_N$.
In the large $N_c$ limit however one can put $m_\Theta\approx m_N$.

Let us consider now the nucleon matrix element of the product of two strangeness-changing
axial currents $j^\alpha_{\mu 5}(x)$ and expand it in intermediate states $|n \rangle$:
\bea\la{Pi1}
&&\Pi^{T=0}(\omega,{\bf q})^i_k=\int\! d^4x\, e^{iq\cdot x}\,
\frac{\varepsilon^{\alpha_1\beta_1}}{\sqrt{2}}\langle N_{\beta_1 k}| j^\dagger_{\nu5\;\alpha_1}(x)j_{\mu 5}^{\alpha_2}
(0)|N^{\beta_2 i}\rangle \frac{\varepsilon_{\alpha_2\beta_2}}{\sqrt{2}}\\
\n
&&=\sum_n \frac{\varepsilon^{\alpha_1\beta_1}}{\sqrt{2}}\langle N_{\beta_1 k}| j^\dagger_{\nu5\;\alpha_1}(0)
|n({\bf q})\rangle\,2\pi\delta\left(\omega-E_n({\bf q})\right)\,
\langle n({\bf q}))|j_{\mu 5}^{\alpha_2}(0)|N^{\beta_2 i}\rangle \frac{\varepsilon_{\alpha_2\beta_2}}{\sqrt{2}}.
\eea
Here $E_n({\bf q})$ is the kinetic energy of the intermediate state. Since the nucleon is infinitely
heavy at large $N_c$ the energy $\omega$ and the 3-momentum ${\bf q}$ are conserved.
We write relativistic equations for the kaon field, however.

The correlation function \ur{Pi1} can be calculated {\it e.g.} in the Skyrme model.
We wish to isolate the kaon pole contribution to the strange axial current,
that is we have to consider $\omega^2\approx {\bf q}^2+m_K^2$ where we temporarily
take the chiral limit, $m_K=0$, for the current to be conserved. In fact, this
requirement can be relaxed. The singular contribution to the current arises from the
asymptotics of the kaon scattering wave \ur{asympt}:
\beq
j^{\alpha}_{\mu 5}(\omega, {\bf q})=iq_\mu\,F_\pi\,K^\alpha(\omega, {\bf q})
=-q_\mu\,F_\pi\,\frac{\sin\delta(\omega)}{\omega^2-{\bf q}^2}\,
\left({\bf q}\cdot\mbox{\boldmath{$\tau$}}\right)^\alpha_{\beta}
{\bf b}^{\dagger\,\beta}(\omega)\,\sqrt{\frac{4\pi}{\omega^3}} .
%\left\{\begin{array}{c} \omega,\quad \mu=0,\\q_i,\quad \mu=i.\end{array}\right.
\la{sax}\eeq
The last factor arises here in accordance with the commutation relation for the
creation-annihilation operators $[{\bf b}^{\dagger\,\beta}(\omega_1){\bf b}_\alpha(\omega_2)]
=2\pi\,\delta^\beta_\alpha\,\delta(\omega_1-\omega_2)$~\cite{CK}.
Substituting \Eq{sax} into \Eq{Pi1} we obtain
\beq
\Pi^{T\!=\!0}_{\mu\nu}\left(\omega, {\bf q}\right)^i_k
=q_\mu q_\nu\,\frac{4\pi F_\pi^2}{\omega^3}\,{\bf q}^2\,\delta^i_k\,
\frac{\sin^2\delta(\omega)}{(\omega^2-{\bf q}^2)^2}\,.
\la{Pi2}\eeq
The correlation function is therefore expressed through the phase shift $\delta(\omega)$!
The conservation of the axial current implies that there is also a contact term in the correlation
function, proportional to $g_{\mu\nu}$; the coefficient in front of it must be exactly the
coefficient in front of $q_\mu q_\nu/q^2$, with the minus sign.

Let us now assume that one of the intermediate states in \Eq{Pi1} is a {\em narrow}
$\Theta^+$ resonance. Then, on the one hand, at $\omega\approx\Delta$ the phase shift $\delta(\omega)$
must exhibit the Breit--Wigner behaviour as it follows from unitarity:
\beq
\sin^2\delta(\omega) = \frac{\Gamma^2/4}{(\omega-\Delta)^2+\frac{\Gamma^2}{4}}\quad
\stackrel{\Gamma\to 0}{\longrightarrow}\quad\frac{\Gamma}{4}2\pi\delta(\omega-\Delta).
\la{phaseT}
\eeq
On the other hand, one can extract the contribution of the $\Theta^+$ intermediate state
using the definition of the matrix elements of the axial current \ur{gandG}.
Taking there the contribution that is singular near the kaon pole
and recalling \Eq{nonrel} we get
\beq
\Pi^{T\!=\!0}_{\mu\nu}\left(\omega, {\bf q}\right)^i_k
=2\pi\delta(\omega-\Delta)\frac{q_\mu q_\nu}{\left(q^2\right)^2}
F_\pi^2 G^2\frac{{\bf q}^2}{4m_N^2}\delta^i_k.
\la{corr-G}
\eeq
We now compare \Eq{corr-G} and \Eqs{Pi2}{phaseT} and immediately obtain the already familiar
equation for the $\Theta^+$ width through the $\Theta KN$ pseudoscalar coupling $G$
({\it cf.} Section 7):
\beq
\Gamma = \frac{G^2(\Theta\to KN)}{4\pi m_N^2}{\bf p}^3
\la{width97}
\eeq
where ${\bf p}$ is the kaon momentum, equal to $\Delta$ in the chiral limit.
We stress that the Born graph for the $KN$ scattering with pseudoscalar Yukawa coupling
arises automatically -- through unitarity -- from the $KN$ scattering phase,
provided it corresponds to a narrow resonance.\\

To conclude, if $\Theta^+$ happens to be a narrow resonance, one can find
its width from the $\Theta KN$ transition axial coupling or, thanks to the Goldberger--Treiman
relation, from the transition pseudoscalar coupling (it is contrary to the recent claim of
Ref.~\cite{WW,Weigel}). This is how the narrow $\Theta^+$
has been first predicted~\cite{DPP-97} and how a more stringent estimate of the width
$\Gamma_\Theta\sim 1\,{\rm MeV}$ has been recently performed~\cite{DP-05,Lorce,LKG}.

\section{Finite-$N_c$ effects in the $\Theta^+$ width}

In any chiral soliton model of baryons, the baryon-baryon-meson coupling can be written
in terms of the rotational coordinates given by the $SU(3)$ matrix R as~\cite{AdkinsNappiWitten}
\beq
L= -ip_i\frac{3G_0}{2m_N}\,\half\,\Tr(R^\dagger\lambda^aR\tau^i)
\la{BBM}\eeq
where $\lambda^a$ is the Gell-Mann matrix for the pseudoscalar meson of flavour $a$, and
$p_i$ is its 3-momentum. The pseudoscalar coupling $G_0$ is directly related to the asymptotics
of the Skyrmion profile function $P(r)\approx r_0^2/r^2\;\;$~\cite{AdkinsNappiWitten} :
\beq
G_0=\frac{8\pi}{3}F_\pi m_Nr_0^2.
\la{G0}\eeq
Note that $G_0={\cal O}(N_c^{\frac{3}{2}})$.
In a generic case there are baryon-baryon-meson couplings other than \ur{BBM},
labeled in Ref.~\cite{DPP-97} by $G_1$ and $G_2$. It is the interplay of these constants
that leads to a small $\Theta^+$ width. In the nonrelativistic limit the combination
of $G_{0,1,2}$ is such that the width goes to zero strictly, however each of the constants
remain finite being then determined solely by valence quarks. Unfortunately, in the Skyrme model
$G_{1,2}$ are altogether absent, related to the fact that there are no valence quarks in the Skyrme model.
For example, $G_2$ is proportional to the fraction of nucleon spin carried
by valence quarks which is known to be exactly zero in the Skyrme model~\cite{BEK}.
Since only the coupling $G_0$ is present in the Skyrme model, we are forced to mimic
the nonrelativistic limit there by taking the size $r_0$ to zero,
which leads to unrealistic parameters. In any chiral model with explicit valence quarks
there are less traumatic ways to obtain a very small $\Theta^+$ width.

In the chiral limit $SU(3)$ symmetry is exact, therefore \Eqs{BBM}{G0} determine also the
leading term in the $\Theta\to KN$ decay width, provided $\Theta^+$ is understood as
an excited rotational state of a nucleon~\cite{DPP-97}. For arbitrary $N_c$ the appropriate
Clebsch--Gordan coefficient has been computed by Prasza\-lowicz~\cite{MichalNc}:
\beq
\Gamma_{\Theta}(N_c)=\frac{3(N_c+1)}{(N_c+3)(N_c+7)}\,\frac{3}{8\pi m_N^2}G_0^2|{\bf p}|^3.
\la{GNc}\eeq
To compare it with the width computed in Section 7 from the imaginary part of the pole in the
kaon-Skyrmion scattering amplitude, one has to take the limit $N_c\to\infty$, as only in this limit
the use of the Callan--Klebanov linearized scattering approach is legal. Using \ur{G0} we find
\beq
\Gamma_{\Theta}(N_c\to\infty)=\frac{8\pi F_\pi^2r_0^4}{N_c}\Delta^3 = {\cal O}(N_c^0),
\la{GNcinfty}\eeq
which coincides exactly with the width obtained in Section 7 by other methods, in particular
from the resonance pole position, where one has to put the coefficient $\gamma\!=\!1$.
To guarantee the validity of this result one has to make sure that the width is small,
$\Gamma\!\ll\!\Delta$, for example, by taking the limit considered in subsection 7.2.
In more realistic models the condition $\Gamma\!\ll\!\Delta$ can be achieved not by taking small
$G_0$ but due to the cancelation of several pseudoscalar coupling $G_{0,1,2}$ as it in fact
must happen in the nonrelativistic limit. Then, as shown from unitarity in Section 8, \Eq{GNc} modified
to incorporate other couplings~\cite{DPP-97,MichalNc} remains valid.

Looking into \Eq{GNc} we can discuss what happens with the width as one goes from
the idealized case of $N_c=\infty$ to the real world with $N_c=3$. Unfortunately, at finite
$N_c$ the whole Skyrmion approach becomes problematic since quantum corrections to the
saddle point are then not small. Quantum corrections to a saddle point in general
and here in particular are of two kinds: coming from zero and nonzero modes. Corrections from nonzero
modes can be viewed as a meson loop in the Skyrmion background. As any other quantum loop
in 4 dimensions, it has a typical additional suppression by $1/(2\pi)$ arising from
the integral over loop momenta $\int d^4p/(2\pi)^4$. We remind the reader that in QED
radiative corrections are not of the order of $\alpha$, the fine structure constant, but
rather $\alpha/(2\pi)\approx 10^{-3}$. Therefore, quantum corrections from nonzero modes are
expected to be of the order of $1/(2\pi N_c)\approx 1/20$ and look as if they can be neglected.
As to zero modes, which are the translations and the rotations of the Skyrmion as a whole,
they do not lead to the additional $1/(2\pi)$ suppression. On the contrary, they lead
to ``kinematical'' factors like the Clebsch--Gordan coefficient in \Eq{GNc}, which bear
huge $1/N_c$ corrections. Hence it is desirable to take rotations into account exactly for any $N_c$.

We are therefore inclined to take \Eq{GNc} at face value for any $N_c$ and claim that
it is the leading effect in accounting for finiteness of $N_c$. At $N_c\!=\!3$ it leads
to the relation
\beq
\Gamma_\Theta(N_c\!=\!3) = \frac{1}{5}\,\Gamma_{\Theta}(N_c\!\to\!\infty).
\la{one-fifth}\eeq
The Clebsch--Gordan coefficient ``1/5'' was actually used in the original paper~\cite{DPP-97}
predicting a narrow pentaquark.
A large suppression of $\Gamma_\Theta$ as compared to its asymptotic value at $N_c\!=\!\infty$
has been also noticed in Ref.~\cite{WW} in another estimate of the finite $N_c$ effects.
Whatever is the width found from the pole position in the kaon-Skyrmion scattering amplitude,
the real $\Theta^+$ width is expected to be at least 5 times less! Estimates for the real-world
$N_c\!=\!3$ in Refs.\cite{DP-05,Lorce,LKG} demonstrate that it can easily by obtained
at the level of 1 MeV or even less, without any fitting parameters.

\section{Conclusions}

The remarkable idea of Skyrme that baryons can be viewed as nonlinear solitons of the
pion field, finds a justification from the modern QCD point of view. However, the concrete realization
of this idea -- the use of the two- and four-derivative Skyrme Lagrangian supplemented by
the four-derivative Wess--Zumino term -- is an oversimplification of reality.
Therefore, the Skyrme model as it is, may work reasonably well for certain baryon observables
but may fail qualitatively for other.

To understand where the Skyrme model fails, one has to keep in mind that the model
implies that the valence quarks are close to the
negative-energy Dirac sea and cannot be separated from the sea that is strongly distorted.
The number of $Q\bar Q$ pairs in a baryon, corresponding to a strongly polarized
sea is ${\cal O}(1)$, times $N_c$. Exotic baryons are then not distinguishable from ordinary ones
as they differ only by one additional $Q \bar Q$ pair as compared to the infinite ${\cal O}(N_c)$
number of pairs already present in the nucleon in that model, hence the exotic decays
are not suppressed. In principle, it does not contradict QCD at strong coupling,
however in reality we know that the octet and decuplet baryons are mainly `made of' $N_c\!=\!3$
constituent quarks with only a small (order of 30\%) admixture of the $N_c\!+\!2\!=\!5$ quark
Fock component. In fact there is an implicit small parameter in baryon physics that may be
called ``relativism'' $\epsilon\ll 1$ such that valence quark velocities are
$v^2/c^2\sim\epsilon$ and the number of $Q\bar Q$ pairs is $\epsilon\,N_c$~\cite{DP-05}.
For observables where the ``nonrelativism'' is essential one expects a qualitative disagreement
with the Skyrme model predictions. For computing such observables it is better, while preserving
the general and correct Skyrme's idea, to use a model that interpolates between the two extremes:
the Skyrme model and the nonrelativistic quark model where there are no antiquarks at all.

Quantization of the $SU(3)$ zero rotational modes of the Skyrmion, whatever is its dynamical realization,
leads to the spectrum of baryons forming a sequence of bands: each band is characterized by ``exoticness'',
{\em i.e.} the number of additional $Q\bar Q$ pairs minimally needed to form a baryon multiplet.
Inside the band, the splittings are ${\cal O}(1/N_c)$ whereas the splittings between bands with increasing
exoticness is ${\cal O}(1)$, see Fig.~5. At large $N_c$ the lowest-mass baryons with nonzero exoticness
(like the $\Theta^+$ baryon) have rotational wave functions corresponding to a small-angle precession.
Therefore, the $\Theta^+$ and other exotic baryons can be, at asymptotically large $N_c$, studied
{\em \`a la} Callan--Klebanov by considering the small oscillations of the kaon field about a Skyrmion,
or the kaon-Skyrmion scattering in the linear order.

This problem has been solved numerically by Klebanov {\em et al.} ~\cite{Klebanov} who have found
that there is no resonance or bound state with the $\Theta^+$ exotic quantum numbers at least in
the large $N_c$ limit, and suggested that it therefore could be an artifact of the rigid rotator
approximation. In this paper, we study this scattering in more detail and come to the opposite conclusion.
While reproducing numerically the phase shifts found in~\cite{Klebanov} we find, both analytically
and numerically, that there is a pole in the complex energy plane, corresponding to a strong $\Theta^+$
resonance which would have definitely revealed itself in $KN$ scattering. Moreover, its origin is
precisely the $SU(3)$ rotational mode. By varying the Skyrme model parameters, we are able to
make $\Theta^+$ as narrow as one likes, as compared to the resonance excitation energy which can be
hold arbitrary. Being arbitrary it nevertheless coincides with the rotational excitation energy.
To understand better the origin of the $\Theta^+$ width, we have computed it in five different
ways yielding the same result. The problem is not the existence of $\Theta^+$ which {\em is}
predicted by the Skyrme model and is a rotational excitation there, but what dynamics makes it narrow.

Although we can deform the parameters of the Skyrme model to make a finite-energy $\Theta^+$ narrow,
they are not natural. It is precisely a problem where the deficiency of the Skyrme model mentioned
above becomes, unfortunately, critical. To get a chance of explaining the narrow width, one needs
a model that interpolates between the Skyrme model and the nonrelativistic quarks models. The narrow
$\Theta^+$ is near the nonrelativistic end of this interpolation. Fortunately, the Chiral Quark Soliton
Model makes the job and indeed estimates of the $\Theta^+$ width there appear naturally with
no parameter fitting at the 1 MeV level. \\

It is exciting and challenging to write this paper at the time when experimental evidence in favour
of the exotic pentaquark $\Theta^+$ is still controversial. We hope that we have waived certain theoretical
prejudice against $\Theta$'s existence and its small width, so it must be there.\\

\noindent
{\bf Acknowledgements}\\

\noindent
We have benefited from discussions with many people but most importantly from conversations
and correspondence with Tom Cohen and Igor Klebanov. We are grateful to Klaus Goeke and
Maxim Polyakov for hospitality at Bochum University where this work has been finalized.
D.D. gratefully acknowledges Mercator Fellowship by the Deutsche Forschungsgemeinschaft.
This work has been supported in part by Russian Government grants RFBR-06-02-16786
and RSGSS-3628.2008.2.

\end{document}